\documentclass[fleqn,usenatbib]{mnras}

\usepackage{newtxtext,newtxmath}
\usepackage[T1]{fontenc}

\DeclareRobustCommand{\VAN}[3]{#2}
\let\VANthebibliography\thebibliography
\def\thebibliography{\DeclareRobustCommand{\VAN}[3]{##3}\VANthebibliography}

\usepackage{graphicx}	% Including figure files
\usepackage{amsmath}	% Advanced maths commands
\usepackage{booktabs}
\usepackage{xcolor}
\usepackage{soul}

\newcommand{\NsAll}{1335}
\newcommand{\NsGrown}{1006}  % M_839 >= 0.9 M_max
\newcommand{\NsGrow}{329}    % M_839 < 0.9 M_max (working subset)

\newcommand{\NsMassiveNow}{48}     % in working subset: M_839 > 7.5 Msun
\newcommand{\ipr}{$\mathrm{I}_{\mathrm{fil}}$}        % iintensity distribution along the longest filament
\newcommand{\vpr}{$\mathrm{V}_{\mathrm{fil}}$}

\newcommand{\rev}[1]{#1}

\newcommand{\lng}[1]{#1}
\newcommand{\dana}[1]{#1}

\usepackage{xparse}
\usepackage[normalem]{ulem}
\definecolor{ppcolor}{RGB}{120,0,120}  % dark magenta / plum
\newcommand{\pp}[1]{#1}

\title[Hub-filament systems and massive stars]{Hub-filament systems and the growth of massive stars: episodic accretion, clustered environments, and projection effects}

\author[D. Makarova et al.]{Dana Makarova,$^{1,2}$
Dana Alina$^{*}$,$^{1,2}$
Paolo Padoan$^{3,4}$,
Baurzhan Kumarioldanov,$^{5}$
Mika Juvela$^{6}$,
Dilda Berdikhan$^{2}$
\newauthor
and Naval Bhadari$^{7}$
\\
$^1$  Physics Department, School of Sciences and Humanities, Nazarbayev University, Kabanbay batyr ave, 53, 010000 Astana, Kazakhstan  \href{mailto:dana.alina@nu.edu.kz}{dana.alina@nu.edu.kz}\\
$^2$ Energetic Cosmos Laboratory, Nazarbayev University, Kabanbay batyr ave, 53, 010000 Astana, Kazakhstan\\
$^3$ Institut de Ci\`{e}ncies del Cosmos (ICCUB), Universitat de Barcelona (UB), c. Mart\'{i} i Franqu\`{e}s, 1, 08028 Barcelona, Spain\\
$^4$ Department of Physics and Astronomy, Dartmouth College, 6127 Wilder Laboratory, Hanover, 03755, NH, USA\\
$^5$  Computer Science Department, School of Engineering and Digital Sciences, Nazarbayev University, Kabanbay batyr ave, 53, 010000 Astana, Kazakhstan \\
$^6$ Department of Physics, P.O. Box 64, FI-00014, University of Helsinki, Finland \\
$^7$ Kavli Institute for Astronomy and Astrophysics, Peking University, 5 Yiheyuan Road, Haidian District, Beijing 100871, China \\
}

\date{Accepted XXX. Received YYY; in original form ZZZ}

\pubyear{\the\year{}}

\begin{document}
\label{firstpage}
\pagerange{\pageref{firstpage}--\pageref{lastpage}}
\maketitle

\begin{abstract}
The processes controlling the early mass growth of future massive stars remain poorly understood, particularly the connection of this growth to star clustering and hub-filament systems (HFSs). This connection is difficult to establish observationally, because projection effects and line-of-sight confusion in position-position-velocity (PPV) data can distort the information about the intrinsic filamentary structure.
To investigate this connection, we used a three-dimensional magnetohydrodynamic (MHD) simulation of star formation, where stars are represented by accreting sink particles. We identify clustered stellar environments, reconstruct time-dependent accretion histories, and investigate the relation between enhanced-accretion episodes and the locations of HFSs. We also use line radiative transfer modeling to produce synthetic molecular-line observations and examine how the same structures appear in projected PPV data. In our simulation, we find that 80\% of future massive stars are associated with clustered environments. Their growth is also highly episodic: typically, about 40\% of the accreted mass is gained during periods of enhanced accretion that occupy only about 10\% of the total growth time. Periods of enhanced accretion occur slightly closer to three-dimensional HFS proxies, suggesting a possible link between HFS morphology and episodic accretion in future massive stars. Overall, our results suggest that the early growth of future massive stars is connected to both their clustered environment and the HFS structure of the surrounding gas, and projection effects must be considered when interpreting HFS in PPV data.
\end{abstract}

\begin{keywords}
Stars: massive –- stars: formation –- ISM: clouds –- methods: numerical -- MHD –- radiative transfer 
\end{keywords}

%%%%%%%%%%%%%%%%%%%%%%%%%%%%%%%%%%%%%%%%%%%%%%%%%%

%%%%%%%%%%%%%%%%% BODY OF PAPER %%%%%%%%%%%%%%%%%%

\section{Introduction}
\label{sec:introduction}
Massive stars are rare, but play a major role in the evolution of their environments. Through radiative and mechanical feedback, they influence the structure and evolution of their host molecular clouds and galaxies \citep{2018ARA&A..56...41M}. They also contribute to the chemical enrichment of the interstellar medium (ISM) through stellar winds and supernova explosions \citep{2003Natur.424..285D, 2021RAA....21...77D}. \pp{However, their earliest stages are difficult to observe, because they are heavily embedded in dense molecular gas for a time that may correspond from a few per cent \citep{2002ARA&A..40...27C,Mottram2011} to a significant fraction \citep{2020ApJ...900...82P,Padoan2026} of their lifetimes.} As a result, the physical processes that control their early mass growth, from parsec-scale clumps \pp{and filaments} to dense cores, remain poorly understood \citep{2020SSRv..216...62R, 2018ARA&A..56...41M}.

Massive stars do not form in isolation. 
\rev{They are commonly associated with dense, structured, and often clustered star-forming regions \citep[e.g.,][]{2003ARA&A..41...57L, 2007ARA&A..45..481Z, 2018ARA&A..56...41M}.}
In these regions, the gas is organised into clumps, cores, and filamentary networks \lng{ \citep[e.g.,][]{stahler2005formation, 2010A&A...518L.102A, 2010A&A...518L.103M, 2025ARA&A..63....1B}.}
Filaments are an important part of this structure, since many prestellar and protostellar cores are observed \lng{to reside in}, suggesting that filaments are closely connected to sites of star formation \lng{\citep[e.g.,][]{2010A&A...518L..92W, 2015A&A...584A..91K, 2016MNRAS.459..342M}.}
In regions of massive and clustered star formation, the larger environment can include several filaments, dense central structures, and embedded young stellar groups or clusters \lng{\citep[e.g.,][]{2009ApJ...700.1609M, 2019A&A...629A..81T, 2022A&A...658A.114K, 2023MNRAS.522.3719L, 2018ARA&A..56...41M}.}
These structures provide the physical environment in which future massive stars assemble their mass.

\rev{The role of the local environment in massive-star formation remains a central question, because different formation scenarios give different importance to the dense core, the surrounding clump, and the larger clustered gas reservoir. 
In turbulent-core or core-accretion models, massive stars grow mainly from massive bound cores embedded in dense, high-pressure clumps \citep{McKeeTan2002,2003ApJ...585..850M}. 
In contrast, competitive-accretion or cluster-scale accretion models give a more direct role to the shared gas reservoir and to the gravitational potential of the forming stellar group \citep{2004MNRAS.349..735B, 2006MNRAS.370..488B, 2014prpl.conf..149T}.
More generally, multi-scale views of massive-star formation emphasize that gas supply may involve a hierarchy of structures, from clouds and clumps to filaments, hubs, and dense cores \citep{2018ARA&A..56...41M, 2019MNRAS.490.3061V}. 
In the inertial-inflow picture, massive-star growth can also be driven by large-scale converging flows, rather than only by a pre-existing massive core or the cluster potential \citep{2020ApJ...900...82P}.
These environmental scales may trace different physical aspects of the formation process, such as immediate accretion from a dense core, access to a larger gas reservoir, the influence of the gravitational potential, or simply the level of local star-forming activity. 
This suggests that the early growth of a future massive star may depend not only on the properties of an individual core, but also on the surrounding stellar group and gas reservoir.}

When several filaments converge toward a dense central region, they can form a hub-filament system (HFS). 
% HFSs are commonly discussed as important sites of clustered and massive star formation, because dense hubs can host young stellar objects, massive cores, and forming clusters \citep{2019A&A...629A..81T, 2022A&A...658A.114K, 2023MNRAS.522.3719L}. 
\lng{HFSs were first discussed observationally by \citet{2009ApJ...700.1609M} as systems where several filaments converge toward a central dense region. Observed HFSs span a wide range of spatial scales, from dense-core scales of $\sim 0.1$ pc to clump and cloud scales of $\sim 1$--$10$ pc, with parsec-scale hubs reported in several massive star-forming regions \citep[e.g.,][]{2022MNRAS.514.6038Z, 2022ApJ...930..169B, 2024MNRAS.527.5895D}. Dense hubs are relevant to clustered and massive-star formation because they concentrate large amounts of gas. Observed HFS hubs can reach column densities of at least $N_{\rm H_2}\sim10^{22}\,{\rm cm}^{-2}$, and in some massive star-forming systems their central hubs approach $N_{\rm H_2}\sim10^{23}\,{\rm cm}^{-2}$ \citep[e.g.,][]{2020A&A...642A..87K, 2024MNRAS.527.4244S}. These values are comparable to the high column densities and surface densities discussed for massive-star formation, including the $\Sigma\sim1\,{\rm g\,cm}^{-2}$ threshold proposed by \citet{2008Natur.451.1082K}. Dense hubs can therefore host massive cores, young stellar objects, and embedded stellar groups \citep[e.g.,][]{2019A&A...629A..81T, 2022A&A...658A.114K, 2023MNRAS.522.3719L}.}
In this picture, filaments may connect the larger \lng{surrounding} gas reservoir to the central hub, where massive stars and clusters can form. Therefore, HFSs provide a useful framework for studying how the morphology of the surrounding dense gas is \lng{significant to} the mass assembly of future massive stars.

\lng{It remains unclear how strongly HFS morphology is connected to the early growth of future massive stars, and whether enhanced accretion is preferentially associated with hub regions in the dense gas.} \lng{This connection is also difficult to interpret in observations, because HFSs are usually studied using projected data, such as continuum maps or position-position-velocity (PPV) cubes.} Projection effects can make unrelated structures appear connected on the sky, while line-of-sight confusion and velocity blending can hide or distort real connections between filaments and hubs \citep{2013ApJ...777..173B, 2019MNRAS.485.4509L}. \lng{Therefore, a projected HFS candidate may not always correspond to a physical three-dimensional HFS.} 

To \lng{address} these issues, we \lng{analyze} a three-dimensional magnetohydrodynamic (MHD) simulation of star formation \lng{together with gas morphology reconstructed from passively advected tracer particles} and synthetic molecular-line observations. \lng{The passive tracer particles follow the gas flow in the simulation and allow us to reconstruct the dense gas structures associated with clustered star formation.
From these structures, we identify junction regions where dense branches converge and use their positions as operational 3D hub proxies. Synthetic observations then allow us to examine how these intrinsic structures appear in projected PPV data.} \lng{Throughout this work, we refer to future massive stars as still-growing sink particles that later reach the adopted massive-star threshold, \(M_{\rm max}\sim7.5\,M_\odot\); the full sample definition is given in Section~3.1.}

\dana{In Section~\ref{sec:methods} we describe the simulations and the radiative transfer modeling used in this work. Section~\ref{sec:results} describes the process of identification of future massive stars \pp{and} the characterization of their environments and present the analysis of the mass accretion. Section~\ref{sec:2d3d} shows the comparison between the three-dimensional morphology of the potential hub-filament systems and their projected two-dimensional observational counterparts. In Section~\ref{sec:discussion} we discuss our results, and \pp{we} summarize our work in Section~\ref{sec:conclusion}}.
%In this work, we first identify still-accreting stars that will later become massive and characterize their clustered environments. We then analyze their time-dependent accretion histories and compare enhanced-accretion episodes with tracer-derived junction regions. Finally, we use synthetic moment maps to connect the intrinsic three-dimensional morphology with its projected observational appearance.

\section{Methods}
\label{sec:methods}
\subsection{MHD simulation description}

\pp{
We use a three-dimensional (3D) magnetohydrodynamic (MHD) simulation of supernova-driven interstellar turbulence, originally introduced by \citet{Padoan_2016} and later extended to higher resolution with self-gravity, tracer particles, and accreting sink particles by \citet{Padoan2017}. The simulation follows the evolution of gas in a cubic region of size $L_{\rm box}=250$ pc, with periodic boundary conditions in all directions. The mean hydrogen number density is $n_{\rm H,0}=5,{\rm cm^{-3}}$, corresponding to a total gas mass $M_{\rm box}\simeq 1.9\times 10^6,M_\odot$. The calculation does not include a galactic gravitational field, vertical stratification, or differential rotation, and should therefore be interpreted as an idealized, high-resolution section of a supernova-driven turbulent ISM.}

\pp{
The initial turbulent state was produced in the calculation of \citet{Padoan_2016}. That run was initialized with zero velocity, uniform temperature $T_0=10^4$ K, and a uniform magnetic field $B_0=4.6\,\mu{\rm G}$. Uniform photoelectric heating was implemented following \citet{Wolfire_1995}, with an efficiency $\epsilon = 0.05$ and a far-ultraviolet radiation field strength $G_0 = 0.6$ in units of the Habing field. Radiative cooling was treated assuming optically thin gas, with atomic cooling dominating above $T \gtrsim 100\;\mathrm{K}$. Molecular cooling and cosmic-ray heating were neglected; instead, the gas temperature was floored at $10\;\mathrm{K}$ at high densities. Photoelectric heating was exponentially tapered above a density of $n_{\rm H} = 200\;\mathrm{cm^{-3}}$ to approximate UV shielding in dense molecular structures. Turbulence was driven by randomly distributed supernova explosions at a rate of $6.25,{\rm Myr^{-1}}$. The original run used a $128^3$ root grid with adaptive mesh refinement and reached a maximum spatial resolution of $\Delta x=0.24$ pc.}

\pp{
The star-forming calculation analyzed here is the continuation described by \citet{Padoan2017}. It was restarted from the $t=45$ Myr snapshot of the original supernova-driven simulation, before the introduction of self-gravity. At restart, the root grid was increased to $512^3$ cells, four adaptive mesh refinement (AMR) levels were used, and the maximum resolution became $\Delta x=0.03$ pc. A total of $2.5\times 10^8$ passively advected tracer particles were initialized, each representing a gas mass of approximately $0.008\,M_\odot$. These tracers record the hydrodynamic variables of the gas and are tagged when they accrete onto sink particles, allowing accretion histories to be reconstructed. }

\pp{
The simulation was then evolved for 10.5 Myr without self-gravity. During this phase, the rate of randomly placed supernovae was reduced by a factor of two, to $3.12\,{\rm Myr^{-1}}$, as the calculation began the transition toward supernova feedback from massive stars formed self-consistently in the simulation. Self-gravity was introduced at $t=55.5$ Myr, together with two additional AMR levels, reaching a maximum resolution of $\Delta x=0.0076$ pc. Sink particles were created when the gas density exceed $n_{\rm H}=10^6\,{\rm cm^{-3}}$ and additional collapse criteria were satisfied, including a local minimum of the gravitational potential, negative velocity divergence, and the absence of another sink particle within the exclusion radius (see \cite{2018ApJ...854...35H} for details). Once formed, sink particles accrete nearby gas and are used here as proxies for individual forming stars. At the resolution of $\Delta x=0.0076$ pc, the stellar initial mass function (IMF) is complete for massive stars ($\gtrsim 7.5\,M_{\odot}$), although lower mass stars are also formed.}

\pp{
We analyze the simulation at a reference time $t_{\rm ref}=15.4$ Myr after the introduction of self-gravity and sink particles, corresponding to an absolute simulation time of approximately 70.9 Myr. At this stage, clustered star formation is well underway and many sink particles are still actively accreting, allowing us to study the relation between the growth of future massive stars and the surrounding dense gas structure in a feedback-regulated turbulent ISM.}

\subsection{Selecting future massive stars and their clustered environments}
\label{sec:selection}

\begin{table}
\centering
\caption{Summary of the stellar sample at the reference time, $t_{\rm ref}=15.4$ Myr after self-gravity was included \dana{in the simulation}.}
\label{tab:stars_839_hier}
\begin{tabular}{lr}
%\hline
%\multicolumn{2}{c}{\textbf{Full simulation at $t=15.4$ Myr}} \\
\hline
Total number of stars & \NsAll \\
\hline
$M_{\rm ref}\ge0.9\,M_{\max}$ (near final mass) & \NsGrown \\
\hline
$M_{\rm ref}<0.9\,M_{\max}$ (still accreting; working subset) & \NsGrow \\
\hline
%\multicolumn{2}{c}{\textbf{Working subset: $M_{\rm ref}<0.9\,M_{\max}$}} \\
%\hline
\hspace{0.5cm} $M_{\rm ref}>7.5\,M_\odot$ (AM stars) & \NsMassiveNow \\
\hspace{0.5cm} $M_{\rm ref}<7.5\,M_\odot$ and $M_{\rm max}>7.5\,M_\odot$ (FM stars) & 59 \\
\hspace{0.5cm} $M_{\rm ref}<7.5\,M_\odot$ and $M_{\rm max}\le7.5\,M_\odot$ (NM stars) & 222 \\
% \hspace{0.5cm} $M_{\rm ref}\le7.5\,M_\odot$ & \NsNonMassiveNow \\
% Expected to gain $\ge1\,M_\odot$ (subset of $M_{\rm ref}\le7.5\,M_\odot$) & \NsGainOne \\
\hline
\end{tabular}
\end{table}

At our reference time \dana{$t_{\rm ref}$}, the simulation contains \NsAll\ stars. Of these, \NsGrown\ stars have reached 90\% or more of their maximum mass \dana{$M_{\rm max}$}. We denote stellar masses measured at \dana{the} reference time as $M_{\rm ref}$. 
Since we focus on regions with ongoing active star formation, we restrict our analysis to stars that are still accreting mass with $M_{\rm ref} < 0.9\,M_{\rm max}$. \dana{Applying this criterion results} in our working subset of \NsGrow\ objects. The \dana{sample} includes stars at different evolutionary stages, spanning a range of masses, ages, and accretion states. 
\lng{
Within this working \dana{sample}, we distinguish three stellar categories. Already massive (AM) stars are stars that are already above the adopted massive-star threshold at the reference time, $M_{\rm ref}>7.5\,M_\odot$. Future massive (FM) stars are stars that are still below this threshold at $t_{\rm ref}$, $M_{\rm ref}<7.5\,M_\odot$, but exceed it later in their evolution, $M_{\rm max}>7.5\,M_\odot$. Non-massive (NM) stars remain below the threshold, with $M_{\rm ref}<7.5\,M_\odot$ and $M_{\rm max}\leq7.5\,M_\odot$. In our sample, these categories contain \NsMassiveNow\ AM stars, 59 FM stars, and 222 NM stars, respectively} \pp{(as mentioned earlier, the IMF from the simulation is incomplete for NM stars)}.
\pp{
We adopt $7.5\,M_\odot$ as an operational threshold for massive stars, close to the conventional lower initial mass for core-collapse supernova progenitors, $M_{\rm SN,min}\simeq 8$--$10\,M_\odot$.
} 
We summarize the adopted selection and the resulting sample sizes in Table~\ref{tab:stars_839_hier}. 
% Thus, FM stars are threshold-crossing objects selected from the still-accreting population.

% In particular, \NsMassiveNow\ stars already have $M_{\rm ref} > 7.5\,M_\odot$, while \NsNonMassiveNow\ remain below $7.5\,M_\odot$. Among the stars below this threshold at $t_{\rm ref}$, 59 later exceed $7.5\,M_\odot$, while the remaining 222 do not.

% \lng{Throughout this work, we define future massive \dana{(FM)}  stars as still-accreting stars that are below the adopted massive-star threshold at the reference time, $M_{\rm ref}<7.5\,M_\odot$, but exceed this threshold later in their evolution, $M_{\rm max}>7.5\,M_\odot$. We use this threshold because, in the underlying simulation, sink particles above $7.5\,M_\odot$ are treated as stars with mass-dependent lifetimes and eventual supernova feedback \citep{2020ApJ...900...82P}. The FM stars classification therefore selects stars that cross the massive-star threshold after the reference time.}
% Thus, the future-massive \dana{(FM)} classification is based on each star's maximum mass, rather than only on its mass at the reference time.} 
% We summarize the adopted selection and the resulting sample sizes in Table~\ref{tab:stars_839_hier}.

\lng{Motivated by the clustered nature of star formation in molecular clouds \citep{2003ARA&A..41...57L}, \pp{we apply the DBSCAN clustering algorithm} \citep{10.5555/3001460.3001507} to the \NsGrow\ still-accreting stars. The resulting clusters therefore trace local associations within the actively growing stellar population at the reference time. In this work, the term cluster refers to a DBSCAN-identified spatial association within the still-accreting stellar population, rather than to a complete census of all stars belonging to a physical stellar cluster.}

% Motivated by the clustered nature of star formation in molecular clouds \citep{2003ARA&A..41...57L}, we identify spatially separated stellar groups in 3D by using density-based spatial clustering of applications with noise (DBSCAN; \citealt{10.5555/3001460.3001507}).
% Stellar groups were identified with DBSCAN using a minimum membership of two stars, \dana{regardless of their mass,} and a neighborhood radius of 
\pp{In DBSCAN,} \lng{we require a minimum membership of two stars and adopt a neighbourhood radius of $\varepsilon = 1.25~\mathrm{pc}$. This value is used as a fiducial parsec-scale linking length that provides a practical definition of compact stellar associations among nearby still-accreting stars at the reference time. It should therefore be interpreted as an operational clustering parameter, rather than as a physical boundary of a stellar cluster.}
% Stellar groups were identified with DBSCAN using a minimum membership of two stars and a neighborhood radius of 
% $\varepsilon = 1.25~\mathrm{pc}$. \lng{DBSCAN was applied to the working subset of \NsGrow\ still-growing stars, including both objects already above and still below the adopted massive-star threshold at the reference time. Thus, the grouping is independent of the current stellar mass within this still-growing subset. The value of $\varepsilon$ is used as an operational parsec-scale linking length for nearby forming stars, motivated by the clustered nature of star formation in molecular clouds and by the parsec-scale structure of cluster-forming HFS environments \citep[e.g.,][]{2003ARA&A..41...57L, 2019A&A...629A..81T}. It is not intended to define a unique physical boundary for a stellar cluster.}

% We identify 45 clusters containing 287 stars, while the remaining 42 stars are classified by DBSCAN as noise and are treated as individually forming stars \dana{at $\mathrm{t_{ref}}$}.
We identify 45 clusters containing 287 stars, while the remaining 42 stars are classified by DBSCAN as noise and are treated as non-clustered \pp{accreting} stars at $t_{\rm ref}$. Among the 45 clusters, 22 contain at least one FM star.
% DBSCAN returns 45 clusters which include 287 stars, while 42 stars that were identified as "noise" can be considered as individually forming stars \dana{at $\mathrm{t_{ref}}$}. 
% Among the 45 clusters, 17 clusters contain only low- or/and intermadiate-mass stars, which means that none of the stars from these clusters reach the massive star threshold. The remaining 28 clusters have at least one star that reaches 7.5 $M_\odot$. 
% in the remaining clusters none of the members reaches the adopted massive-star threshold, $M_{\rm max}>7.5\,M_\odot$. 
% The massive stars are accreting at $\mathrm{t_{ref}}$ in \dana{22 groups.} 
% \st{contain future massive stars which at time 15.4 Myr after the self-gravity was included are below 7.5 $M_\odot$ but reaching it later at time}}. 
% For the unclustered stars, 29 are low/intermediate-mass, 1 is already massive at the reference time, and 12 are future massive. So at \dana{\st{this epoch} $\mathrm{t_{ref}}$}, most future massive stars are found in clustered environments (47 out of 59, $\sim 80\%$), while the remaining 12 objects are classified as unclustered. 
\lng{Among the 42 stars classified as noise by DBSCAN, 29 are NM stars, one is an AM star, and 12 are FM stars. Therefore, at $t_{\rm ref}$, most FM stars are associated with DBSCAN-identified clusters: 47 out of 59 objects, or approximately 80\%, are clustered, while the remaining 12 are non-clustered.} \dana{To test whether these unclustered objects represent a long-lived isolated population, we reevaluate their group membership at earlier and later times. We find that 11 objects later become part of a stellar group, on average $\sim 0.72$ Myr later. The only object without cluster membership after $t_{\rm ref}$ was clustered at earlier times, but remains classified as unclustered from $t_{\rm ref}$ onward. Therefore, the unclustered category at the reference time should be interpreted as time dependent rather than a distinct isolated formation mode.}

\begin{figure}
    \centering
    \includegraphics[width=0.48\textwidth]{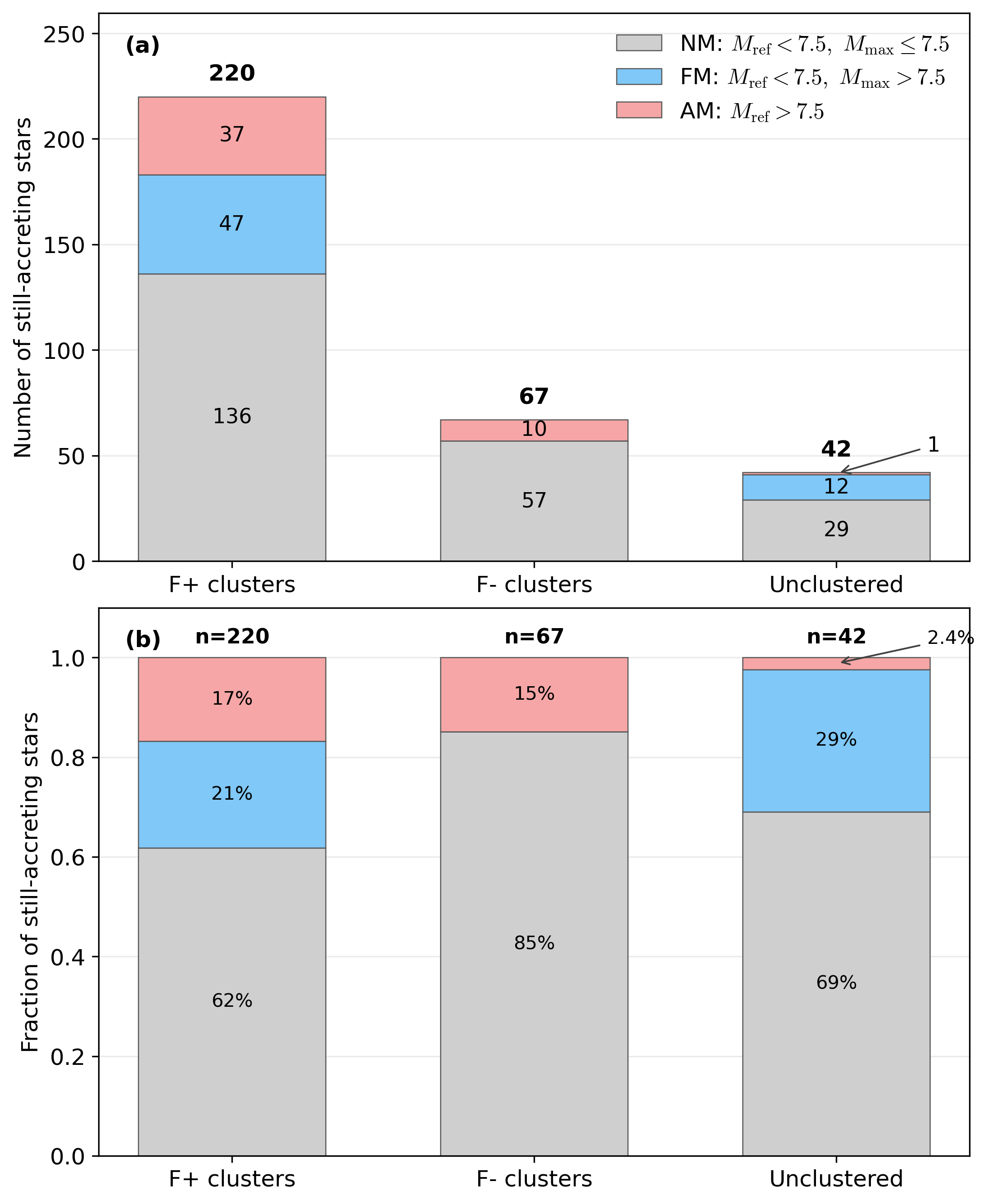}
    \caption{Composition of the DBSCAN sample at the reference time. The stars are grouped according to their DBSCAN environment: F+ clusters contain at least one future massive (FM) star, F- clusters contain no FM stars, and unclustered stars are objects classified as DBSCAN noise. The stacked bars show three stellar classes defined using the stellar mass at the reference time, $M_{\rm ref}$, and the maximum mass reached over the full evolution, $M_{\max}$. NM stars are non-massive stars with $M_{\rm ref}<7.5\,M_\odot$ and $M_{\max}\leq7.5\,M_\odot$; FM stars have $M_{\rm ref}<7.5\,M_\odot$ and $M_{\max}>7.5\,M_\odot$; AM stars are already massive at the reference time, with $M_{\rm ref}\geq7.5\,M_\odot$. Panel (a) shows the absolute number of stars in each category, with the total number shown above each bar. Panel (b) shows the corresponding fractional composition, with the same totals indicated above the bars.
    }
    \label{fig:cluster_statistics}
\end{figure}

% \lng{To quantify the environments in which FM stars are forming, we divide the DBSCAN clusters into two classes according to whether they contain FM stars. Clusters with FM stars are defined as clusters that contain at least one still-growing star with $M_{\rm ref}<7.5\,M_\odot$ and $M_{\rm max}>7.5\,M_\odot$. Clusters without FM stars contain no stars that are below $7.5\,M_\odot$ at $t_{\rm ref}$ and cross this threshold later.} 
% Thus, the remaining clusters are classified as clusters without future massive stars. 
% This definition should be distinguished from the presence of already massive stars at the reference time. The future-massive classification follows the subsequent evolution of still-growing stars, whereas already massive stars are those with $M_{\rm ref}>7.5\,M_\odot$ at the reference time. Therefore, already massive stars can be present in both cluster groups. Among the 22 \lng{clusters with future massive stars}, 5 already host massive stars at the reference time. Among the 23 clusters without future massive stars, 6 also contain already massive stars. Overall, the future-massive group contains 37 already massive stars, while the non-future-massive group contains 10 already massive stars.
\lng{To quantify the environments in which FM stars are forming, we divide the DBSCAN clusters into two classes according to whether they contain FM stars \pp{or not}. We define F+ clusters as clusters that contain at least one FM star, and F- clusters as clusters that contain no FM stars. With this definition, the sample contains 22 F+ clusters and 23 F- clusters. The F+/F- classification refers only to the presence or absence of FM stars and is therefore distinct from the presence of AM stars at the reference time. AM stars can be present in both cluster classes. In our sample, all 22 F+ clusters also contain at least one AM star at $t_{\rm ref}$. Conversely, six F- clusters contain AM stars but no FM stars.} 
% \dana{Here you can add the figure with the clusters content. }
\lng{The F+/F- classification is applied only to DBSCAN clusters. For completeness, we also keep the DBSCAN unclustered population as a separate category in the following summary figure. Figure~\ref{fig:cluster_statistics} shows the absolute and fractional contributions of NM, FM, and AM stars in F+ clusters, F- clusters, and the DBSCAN unclustered population.}

\subsection{Radiative transfer modeling}
\label{sec:loc}

\pp{We carried out radiative transfer calculations to generate molecular emission lines from the simulation using the LOC code \citep{LOC_2020}}. LOC (Line radiative transfer using Open Computing language) is a family of efficient radiative transfer \dana{codes for modeling molecular line emission optimized for execution on GPUs}. 
%It allows us to perform calculations on GPU. 
%Specifically w
We used \pp{the} three-dimensional version of LOC \pp{that} supports hierarchical octree grids and therefore is well-suited for handling the AMR-format data \pp{of the RAMSES code \citep{Teyssier2002} used for the MHD simulation.}

For the \pp{purpose} of the radiative transfer modeling, molecular \pp{abundances,} \pp{not computed in the simulation, were set} to a constant value. The temperature of the gas \pp{was} assumed to be uniform with a value of $T = 15\;\mathrm{K}$, representative of observed filamentary molecular clouds, whose central dust temperatures are typically in the range $11-15\mathrm{K}$ \citep{2011A&A...529L...6A}. Most of the recently observed HFSs are located at the distance of $\gtrsim 1000\;\mathrm{pc}$ \citep{2020A&A...642A..87K, Mookerjea_2023}, so we placed the simulation at the distance of $1000\;\mathrm{pc}$ to \pp{represent} molecular clouds at intermediate distances.

The molecular transitions $^{12}\mathrm{CO}(1$-$0)$, $\mathrm{C}^{18}\mathrm{O}(1$--$0)$ and $^{13}\mathrm{CO}(1$--$0)$ are commonly used in the investigation of HFSs \citep{Dewangan_2020, 2022MNRAS.514.6038Z}. Therefore, we modeled spectral lines of the CO isotopes focusing mainly on $^{13}\mathrm{CO}$ with an assumed fractional abundance of $[\mathrm{^{13}CO}]/[\mathrm{H_2}]=10^{-6}$. We used 500 velocity channels covering a total bandwidth of 100 km/s resulting in a velocity resolution of 0.2 km/s. This spectral resolution is comparable to that of large-scale Galactic $^{13}\mathrm{CO}(1$--$0)$ surveys such as the Galactic Ring Survey (0.21 km/s \dana{according to} \cite{2006ApJS..163..145J}).

%\subsection{Producing synthetic PPV maps}

%\lng{For each selected cluster, the analysis is performed on a local synthetic cube extracted around the cluster. The local cube is centered on the geometric centre of the selected region and is chosen to include the cluster members and the associated gas reservoir; the radiative-transfer setup is described in Section~\ref{sec:loc}.}

For each cluster, we calculated its geometrical centre, rather than a mass-weighted centre, to ensure that the selected field enclosed all associated stars while avoiding a positional bias towards the most massive members. We then extracted a $~\mathrm{4.88 pc \times 4.88 pc}$ area centered on each cluster, while keeping the full 250$~\mathrm{pc}$ along the line of sight. Radiative-transfer calculations with LOC were performed through this complete line-of-sight volume for the selected CO isotopologues, producing position–position–velocity (PPV) cubes (T(x,y,v)).The integrated-intensity (moment 0) and intensity-weighted velocity (moment 1) maps were calculated from the PPV cubes as

\begin{equation}
M_0(x,y) = \int T(x,y,v) dv ,
\end{equation}

\begin{equation}
M_1(x,y) =
\frac{\int vT(x,y,v) dv}
{\int T(x,y,v) dv} .
\end{equation}

\lng{Here, $M_0(x,y)$ is the integrated intensity at the sky position $(x,y)$, $M_1(x,y)$ is the intensity-weighted mean line-of-sight velocity, $T(x,y,v)$ is the brightness temperature at position $(x,y)$ and velocity $v$, and $dv$ is the velocity-channel width.}
We treat projected maps along the $x$, $y$, and $z$ directions as three separate lines of sight toward the same local volume. This allows us to test how the apparent filamentary morphology and projected kinematic structure depend on viewing orientation, and to identify features that remain similar across different projections.

\subsection{Identifying hub candidates in PPV data}
\label{sec:2dhubs}
\dana{To identify potential HFSs in the synthetic observation data, we follow the observational approach used by \citet{2022MNRAS.514.6038Z}, who analysed molecular-line moment maps of HFS and measured intensity and velocity variations along the longest filaments.}

\lng{In our work, for each viewing direction, we use moment maps derived from the synthetic $^{13}{\rm CO}$ PPV spectral cubes.}
%: the integrated-intensity (moment-0) map and the moment-1 map, which gives the intensity-weighted velocity centroid.} 
% For each viewing direction, the analysis is based on moment maps \dana{(integrated intensity and velocity centroid)} derived from the synthetic \pp{PPV} spectral cube.
% \dana{Here we adopted the approach from ..., which is a rather commonly used in HFS studeies (give references)} 
 \lng{Before extracting filaments, we construct an initial mask from the smoothed moment-0 map, using the 75th percentile of the smoothed integrated intensity as the threshold. This mask isolates the coherent high-emission structure and removes low-level diffuse emission. The filamentary network is then identified on the masked moment-0 map with FilFinder \citep{2015MNRAS.452.3435K}. We choose the FilFinder setup to trace coherent filamentary emission on sub-parsec scales while suppressing small isolated structures and holes in the mask. In physical units, the adopted parameters correspond to a beam width of $0.015\,{\rm pc}$, a smoothing scale of $0.038\,{\rm pc}$, and an adaptive-thresholding window of $0.152\,{\rm pc}$. We retain structures with areas larger than $0.029\,{\rm pc}^2$, fill holes smaller than $0.0029\,{\rm pc}^2$, and prune terminal spurs shorter than $0.19\,{\rm pc}$. We retain the longest connected filament as the main projected filament and use it as the path along which projected quantities are measured. We extract one-dimensional profiles of the integrated intensity and velocity centroid as functions of position along the filament skeleton. 
In the following, we refer to these quantities as the integrated-intensity profile and the velocity-centroid profile along the main projected filament. We define 2D hub candidates as high-connectivity regions in the projected FilFinder filaments, corresponding to locations where three or more skeleton segments connect. 
%We then compare their positions with the projected positions of the 3D hub proxies identified in Section~\ref{sec:3dhubs}.
}

% \pp{Comment: should we start already in this paragraph to use position-intensity (PI) and position-velocity (PV) profiles as in the following paragraph? If not, maybe we should keep calling them integrated-intensity profiles and velocity profiles also in the following paragraph. It's not good to use different terms for the same thing.} 

\section{Results}
\label{sec:results}

\subsection{Instantaneous accretion in clusters}
%Stellar content and instantaneous accretion in clusters

\begin{figure}
    \centering
    \includegraphics[width=0.48\textwidth]{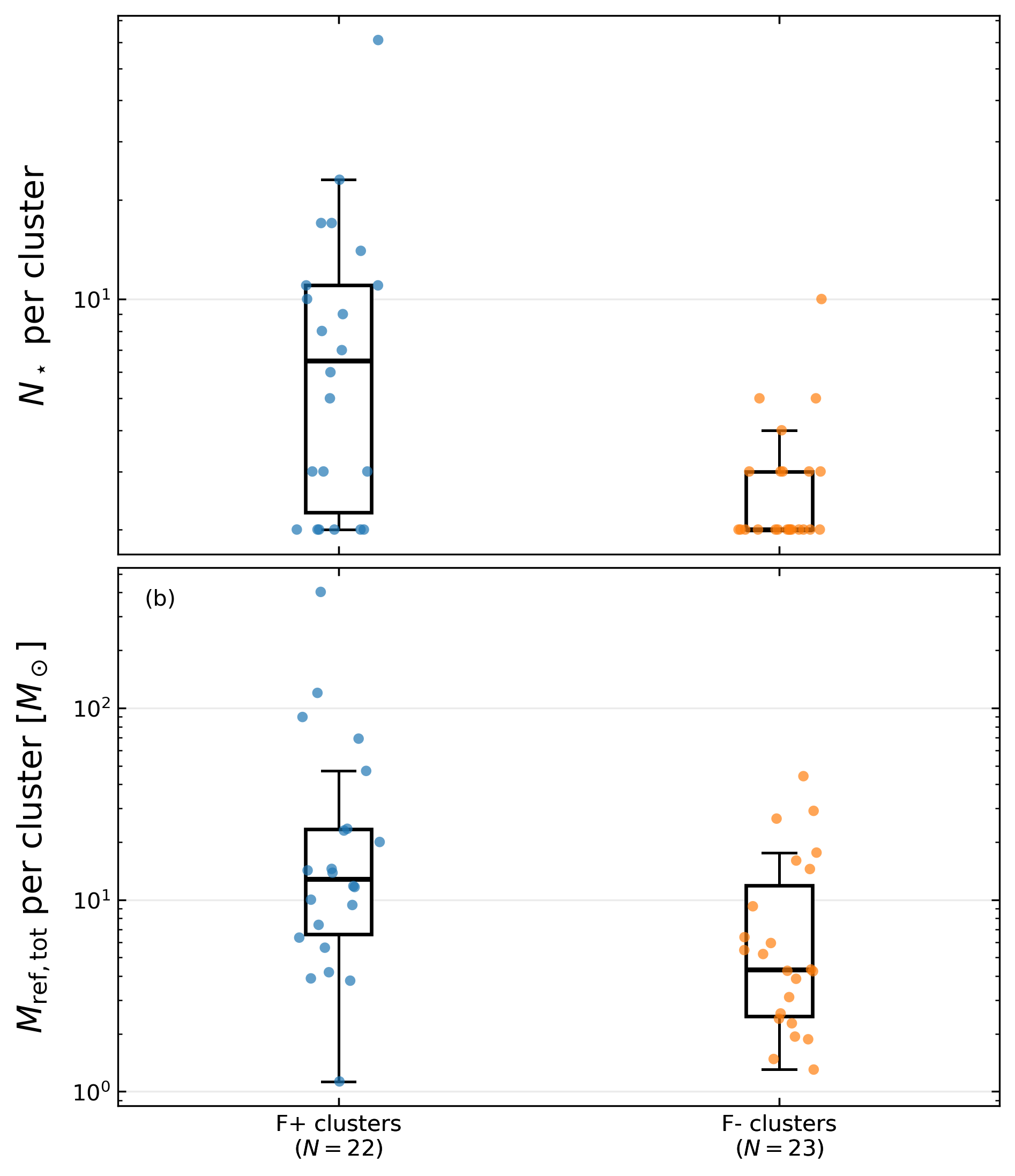}
    \caption{Stellar content of F+ (with future massive stars) and F- (without future massive stars) clusters at the reference time. 
    The upper panel shows the number of stars per cluster, $N_\star$, while the lower panel shows the total stellar mass per cluster, $M_{\rm ref,tot}$. Both panels are shown with logarithmic y-axes.
    Both quantities are computed for the DBSCAN cluster members within the working sample of \NsGrow\ still-accreting stars. 
    Each point represents one DBSCAN cluster; boxes show the interquartile (25th–75th percentiles) range, horizontal lines mark the medians, and whiskers extend to the most extreme data points within $1.5$ times the interquartile range.
    % \pp{Comment: maybe the median line should be blue for the F+ clusters if it is orange for the F- ones?} \pp{Comment: the Y-title of the bottom plot should be $M_{\rm ref,tot}$, and with latex math fonts possibly. The Y-title of the top plot should be $N_\star$. I also wonder if using F+ and F- clusters, as in the main text (and as in Figure 2), may be more clear than calling them "future-massive" and "non-future-massive", which sounds a bit awkward.}
    }
    \label{fig:q1_boxplot_column}
\end{figure}

We characterize each cluster by three stellar-content metrics measured at the reference time: the number of stars, $N_\star$, the total stellar mass, $M_{\rm tot,ref}=\sum M_{\rm ref}$, and the maximum stellar mass, $M_{\rm max,ref}=\max(M_{\rm ref})$.
These quantities are computed using only the still-accreting working sample of \NsGrow\ stars. Table~\ref{tab:cluster_census_839} summarizes the distributions of these metrics for the F+ and F- cluster samples using the median, interquartile range (25th–75th percentiles), and the minimum and maximum values.
% \dana{In what follows, these metrics are referred to as \st{All}} richness metrics, and are computed for the working subset of $\NsGrow$ stars. % that are still accreting mass.
% We find that clusters with future massive stars are systematically richer in number of stars than clusters without massive stars in future.
\lng{We find that F+ clusters typically contain more still-accreting stars than F- clusters.
% When we compare \dana{these two groups of} clusters \dana{\st{with future massive stars to those without,}} we see that the former typically \dana{\st{have more forming}} form more stars. 
%In contrast, the clusters without future massive stars are significantly poorer on median. %% I remived because it is obvious
% Clusters with future massive stars also have higher total stellar mass than clusters without future massive stars.
They also have higher total stellar masses at the reference time.
% The high-mass tail of the cluster mass distribution is dominated by clusters with massive stars in future (7 of the 10 most massive clusters by $M_{\rm tot,ref}$). Figure~\ref{fig:q1_boxplot_column} visualizes these distributions and highlights the extended high-mass tail among \lng{clusters with future massive stars}.
The high-mass tail of the cluster mass distribution is dominated by F+ clusters: 7 of the 10 most massive clusters by $M_{\rm tot,ref}$ belong to the F+ sample. Figure~\ref{fig:q1_boxplot_column} visualizes these distributions and highlights the extended high-mass tail among F+ clusters.}

\begin{table*}
\centering
\caption{Summary of stellar-content metrics at the reference time for the two DBSCAN cluster classes. F+ clusters contain at least one future massive star, while F- clusters contain no future massive stars. Values shown are the median, the interquartile range (25th-75th percentiles), and the minimum and maximum values. \pp{Mass values are in $M_{\odot}$.}
% The last row reports DBSCAN noise objects aggregated as a single sample (\dana{stars not associated with clusters at $\mathrm{t_{ref}}$}).
}
\label{tab:cluster_census_839}
\begin{tabular}{lccccc}
\toprule
Group & $N_{\rm cl}$ & $N_\star$ per cluster & $M_{\rm tot,ref}$ per cluster & $M_{\rm max,ref}$ per cluster \\
\midrule
\lng{F+ clusters} & 22  &
6.5 [2.2, 11.0] (2--61) &
12.8 [6.6, 23.3] (1.1--402.9) &
6.0 [3.2, 7.2] (0.7--20.7) \\
\lng{F- clusters} & 23  &
2.0 [2.0, 3.0] (2--10) &
4.3 [2.5, 11.9] (1.3--44.1) &
2.4 [1.3, 5.8] (0.9--23.1) \\
% Noise / isolated stars & --  &
% 42 (total) &
% 114.0 (total) &
% 17.0 \\
\bottomrule
\end{tabular}
\end{table*}

% To quantify the strength of these differences, we complement the descriptive comparison with non-parametric statistical tests. 
Because the cluster properties show uneven distributions with an extended high-value tail, \lng{we compare the F+ and F- samples using a two-sided Mann--Whitney U test \citep{MannWhitney1947}.} This rank-based test evaluates whether values in one sample tend to be systematically larger or smaller than in the other. We find a significant difference in the number of stars per cluster, $N_\star$ ($p=0.0019$), and in total cluster mass, using $\log_{10} M_{\rm tot,ref}$ ($p=0.0106$).
Since the p-value quantifies the evidence for a difference, but it does not measure the physical size of the effect, we also report Cliff's delta ($\delta$), a non-parametric effect size that quantifies the degree of separation between two distributions \citep{Macbeth2011}. We obtain $\delta=0.52$ for $N_\star$ and $\delta=0.45$ for $\log_{10} M_{\rm tot,ref}$, indicating a clear shift toward larger values in the F+ sample. \lng{These tests therefore support the descriptive result that F+ clusters contain more still-accreting stars and have higher total stellar masses than F- clusters.} 

% Building on the results above, a related question is whether \lng{clusters with future massive stars} with higher richness also tend to have stars that gain mass more rapidly at this evolutionary stage.
Building on the results above, a related question is whether the larger stellar content of F+ clusters is also associated with stronger current accretion activity.
% \dana{\st{We address this by contrasting the two cluster populations introduced above.}} 
To quantify ongoing growth, we estimate the instantaneous mass accretion rate for each star using a central finite difference,
\begin{equation}
\dot{M}_{\rm ref} \approx
\frac{M(t_{\rm ref}+\Delta t)-M(t_{\rm ref}-\Delta t)}
{2\Delta t},
\end{equation}
where $\Delta t = 29.5$ kyr is the time spacing between consecutive simulation outputs. 

% For the normalized metric, we use only $\dot{M}_{\rm ref}>0$ to keep the ratio well defined. 
% The total cluster accretion rate is defined as $\dot{M}_{\rm cl,sum}=\sum_{i\in{\rm cluster}}\dot{M}_{i,\rm ref}$  which captures the overall mass growth of the custers.
\lng{The total cluster accretion rate is defined as $\dot{M}_{\rm cl,sum}=\sum_{i\in{\rm cluster}}\dot{M}_{i,\rm ref}$, where the sum is taken over all still-accreting stars assigned to the cluster.
This quantity captures the total instantaneous mass growth of the cluster.}
To reduce the direct dependence on the number of stars in a cluster, we also compute the typical accretion level per star in each cluster using $\dot{M}_{\rm cl,med}=\mathrm{median}_{i\in{\rm cluster}}(\dot{M}_{i,\rm ref})$.
\lng{Comparing F+ and F- clusters, we find that F+ clusters have significantly higher total accretion rates.}
The median value is $\dot{M}_{\rm cl,sum}=1.13\times10^{-5}\,M_\odot\,{\rm yr}^{-1}$ for F+ clusters, compared to $3.84\times10^{-6}\,M_\odot\,{\rm yr}^{-1}$ for F- clusters. A two-sided Mann--Whitney U test gives $p=0.0033$, with Cliff's delta $\delta=0.51$, indicating a clear shift toward higher $\dot{M}_{\rm cl,sum}$ in the F+ sample.

% The median $\dot{M}_{\rm cl,sum} = 1.13\times10^{-5}\,M_\odot{\rm yr}^{-1}$ for clusters with future massive star versus $3.84\times10^{-6}M_\odot{\rm yr}^{-1}$ for clusters without future massive stars, with a two-sided Mann–Whitney test yielding $p=0.0033$ and a large effect size $\delta=0.51$, indicating a substantial shift toward higher $\dot{M}_{\rm cl,sum}$ in the future-massive population. 
% In contrast, the distribution of the $\dot{M}_{\rm cl,med}$ are similar between the two groups, the medians are $1.3\times10^{-6}$ and $1.13\times10^{-6}M_\odot{\rm yr}^{-1}$ for future massive and non-future massive clusters respectively, and the difference is not significant ($p=0.67$, $\delta=0.08$). 

\begin{figure}
    \centering
    \includegraphics[width=0.48\textwidth]{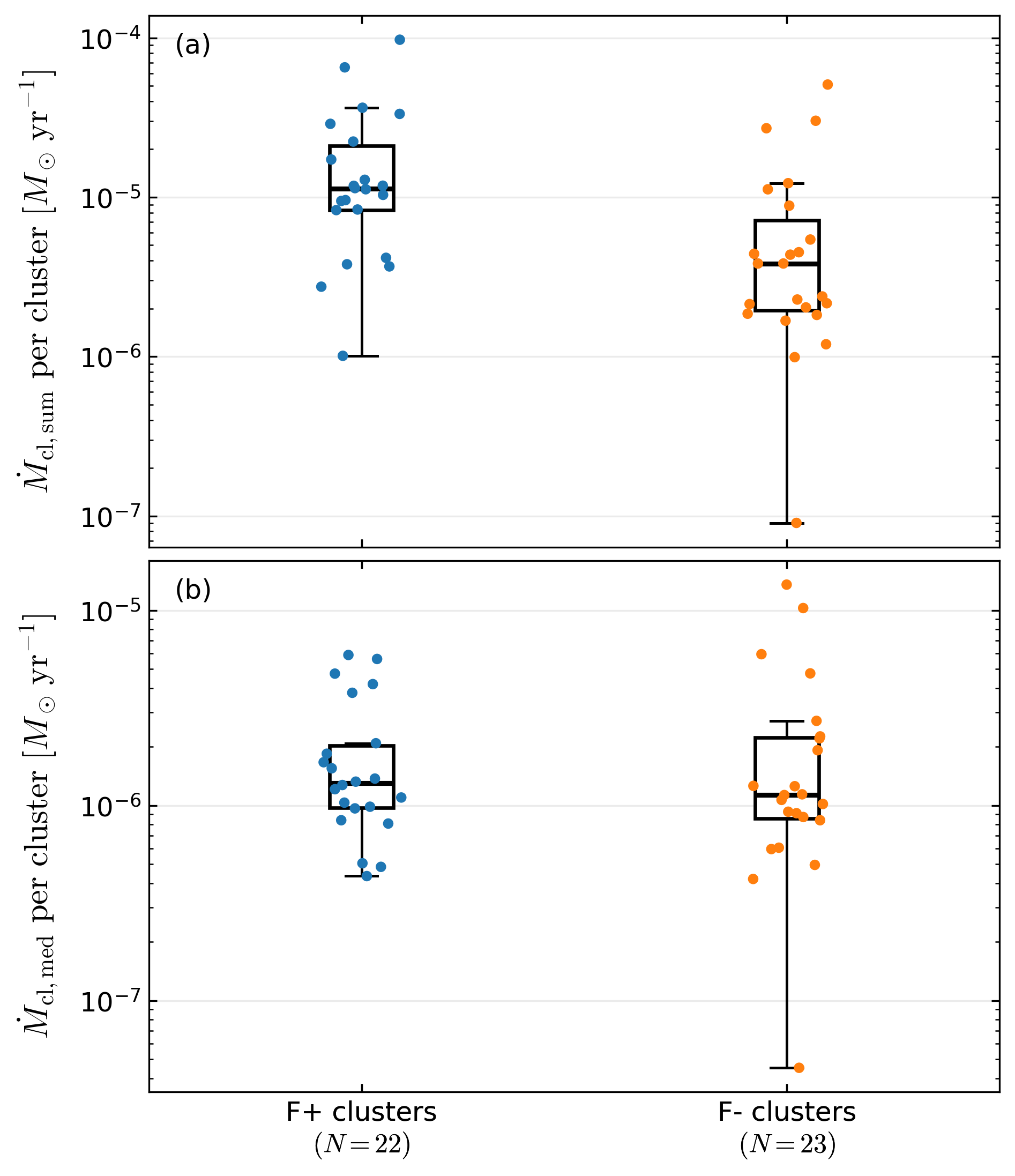}
    \caption{
    % (a) Total cluster accretion rate, $\dot{M}_{\rm cl,sum}=\sum_{i\in{\rm cluster}}\dot{M}_{i,\rm ref}$, for clusters that host at least one future massive star and for clusters without future massive stars. (b) Typical per-star accretion level in each cluster, $\dot{M}_{\rm cl,med}=\mathrm{median}_{i\in{\rm cluster}}(\dot{M}_{i,\rm ref})$. Each point represents one cluster and boxplots show the median and interquartile range (outliers are not shown). Only stars with $\dot{M}_{\rm ref}>0$ are included when computing the cluster metrics 
    % \pp{(still-accreting stars may temporarily have no accretion at the reference time).} 
    % \pp{Comment/question: I wonder how many $\dot{M}_{\rm ref}=0$ stars you find, and how they are distributed among F+ and F- clusters --mainly a curiosity, not necessarily something to report.} Comment: not so many. Total stars with dotM_ref = 0: 6 / 329 Clustered stars with dotM_ref = 0: 2 / 287 Unclustered stars with dotM_ref = 0: 4 / 42
    % The y-axes use a symmetric log scale to visualize the broad dynamic range. 
    % \pp{Comment: (a) and (bb) are missing from the plots. Even you put them in the panels or refer to the panels as top and bottom.}
    Cluster-scale accretion rates at the reference time for $F+$ and $F-$ clusters. Panel (a) shows the total cluster accretion rate, $\dot{M}_{\rm cl,sum}=\sum_{i\in{\rm cluster}}\dot{M}_{i,\rm ref}$. Panel (b) shows the typical per-star accretion level in each cluster, $\dot{M}_{\rm cl,med}=\mathrm{median}_{i\in{\rm cluster}}(\dot{M}_{i,\rm ref})$. Both quantities are computed using cluster members with positive instantaneous accretion rates, $\dot{M}_{i,\rm ref}>0$ (still-accreting stars may temporarily have no accretion at the reference time). Each point represents one DBSCAN cluster; boxes show the interquartile ((25th–75th percentiles) range, horizontal lines mark the medians, and whiskers extend to the most extreme data points within $1.5$ times the interquartile range. Both panels are shown with logarithmic y-axes.
}
    \label{fig:q2a_boxplot_column}
\end{figure}

In contrast, the distributions of $\dot{M}_{\rm cl,med}$ are similar for the two cluster classes. The median values are $1.3\times10^{-6}$ and $1.13\times10^{-6}\,M_\odot\,{\rm yr}^{-1}$ for F+ and F- clusters, respectively, and the difference is not significant ($p=0.67$, $\delta=0.08$). 
\lng{Taken together, these two metrics indicate that the higher total accretion rate of F+ clusters mainly reflects their larger number of still-accreting stars, rather than a systematically higher median accretion rate per cluster member.} The difference in $\dot{M}_{\rm cl,sum}$ and the similarity in $\dot{M}_{\rm cl,med}$ are illustrated in Figure~\ref{fig:q2a_boxplot_column}. 
% We therefore next examine the accretion rates of individual FM stars directly.

% Taking into account both metrics, this indicates that enhanced growth on the cluster level in clusters with future massive star mainly reflects their larger forming population, rather than a systematically higher accretion rate for a typical star.
% The difference in $\dot M_{\rm cl,sum}$ and the similarity in $\dot M_{\rm cl,med}$ are illustrated in Figure ~\ref{fig:q2a_boxplot_column}.

While the comparison of $\dot{M}_{\rm cl,sum}$ and $\dot{M}_{\rm cl,med}$ describes the overall accretion activity of each cluster, it does not directly show whether the FM stars themselves are already accreting faster. This distinction is important because the median per-star accretion rate, $\dot{M}_{\rm cl,med}$, is similar in F+ and F- clusters. We therefore next examine accretion at the star level. The FM stars are compared to NM stars, which are also below the threshold at $t_{\rm ref}$ but never cross it later. This choice avoids mixing in AM stars and isolates differences related to the later evolutionary outcome rather than to the current mass range.
% The \lng{FM sample} consists of stars with $M_{\rm ref}<7.5\,M_\odot$ and $M_{\rm max}>7.5\,M_\odot$. As a control sample, we use NM stars, which are also below the threshold at the reference time but never cross it later, $M_{\rm ref}<7.5\,M_\odot$ and $M_{\rm max}\leq7.5\,M_\odot$. This choice avoids mixing in AM stars and isolates differences related to the later evolutionary outcome rather than to the current mass range.
% While these \dana{\st{cluster scale}} results describe the overall growth activity \dana{\st{of the environment}}, they do not directly address whether specific stars are already growing faster. In particular, future massive stars may represent a special subset whose current accretion is higher even if the typical cluster member is not. Therefore, we next examine accretion at the star level and focus on stars that are still sub massive at \dana{$\mathrm{t_{ref}}$} \st{the reference time}. Since we define future massive stars as objects with $M_{\rm ref}<7.5\,M_\odot$ and $M_{\max}>7.5\,M_\odot$, for the control sample we use accreting \dana{\st{ sub-massive} stars \st{at the same time $M_{\rm ref}<7.5\,M_\odot$} that will never cross the threshold of $7.5\,M_\odot$}. This choice avoids mixing in stars that are already massive at the reference time and isolates differences related to the future outcome rather than the current mass range. 
% We compare the $\dot{M}$ distributions of these two sub massive samples across the full working subset. 
% At the reference time, future massive stars already show higher instantaneous accretion. 
Among stars that are below the massive-star threshold at $t_{\rm ref}$, \lng{FM stars already show enhanced instantaneous accretion. The median accretion rate is $\dot{M}_{\rm ref}=1.99\times10^{-6}\,M_\odot\,{\rm yr}^{-1}$ for the FM sample ($N=58$), while the NM sample has a median value of $7.43\times10^{-7}\,M_\odot\,{\rm yr}^{-1}$ ($N=217$).} The difference is highly significant with $p=6.4\times10^{-7}$ and a clear separation between the distributions $\delta=0.43$. This suggests that enhanced mass growth is already present in the stars that will later become massive, even before they cross the $7.5M_\odot$ threshold.

\begin{figure}
    \centering
    \includegraphics[width=0.48\textwidth]{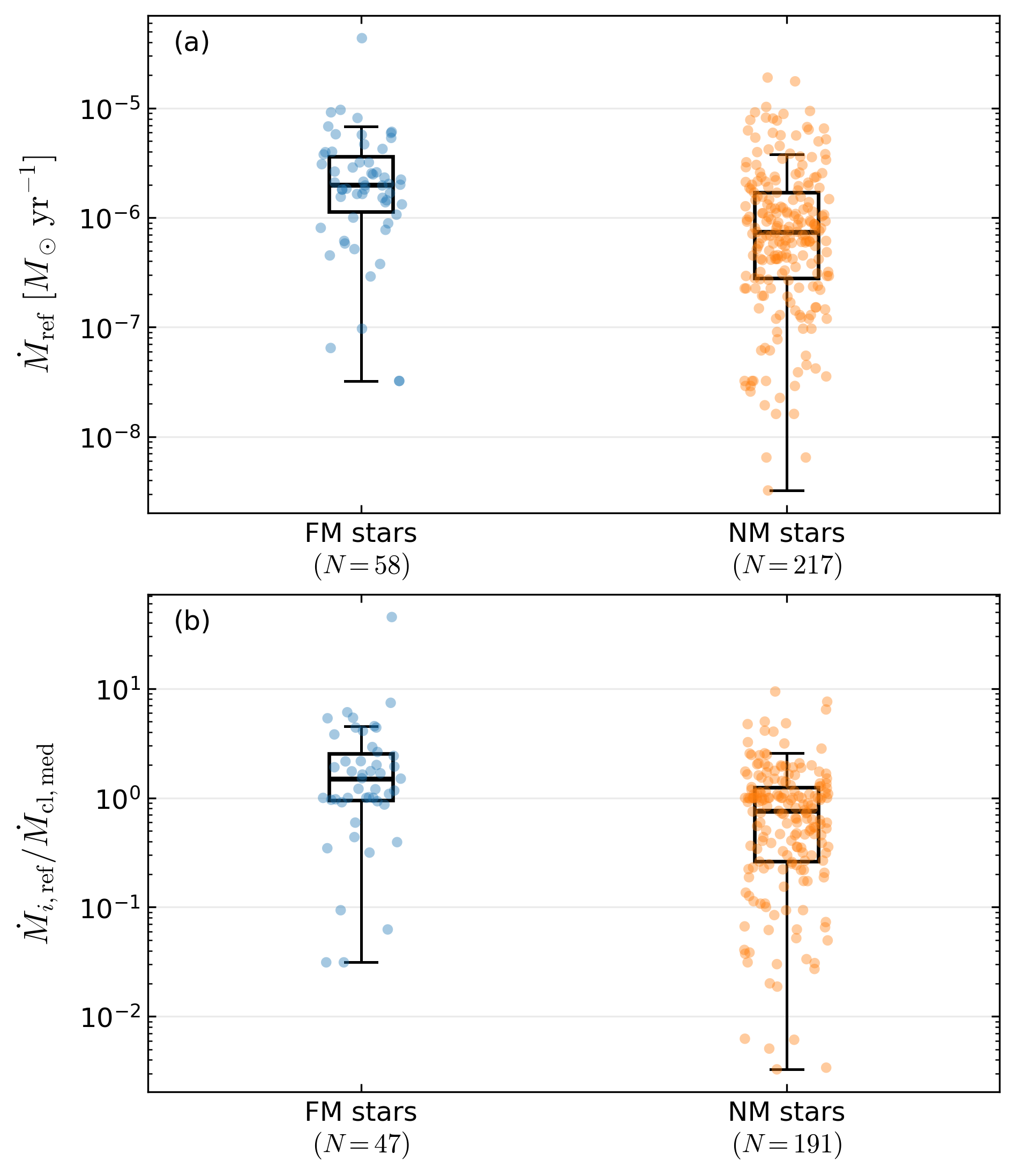}
    \caption{
    % (a) Global comparison of the instantaneous accretion rate $\dot{M}_{\rm ref}$ for future massive stars ($M_{\rm ref}<7.5\,M_\odot$ and $M_{\max}>7.5\,M_\odot$) and a sub-massive control sample ($M_{\rm ref}<7.5\,M_\odot$ and $M_{\max}\le 7.5\,M_\odot$). Only stars with $\dot{M}_{\rm ref}>0$ are included. (b) Within-cluster comparison using the relative accretion rate $\dot{M}_{\rm rel}=\dot{M}_{\rm ref}/\mathrm{median}(\dot{M}_{\rm ref})_{\rm cluster}$, which controls for the typical accretion level of each host cluster. Boxplots show the median and interquartile range; points show individual stars (jittered horizontally). The y-axes use a \ppst{symmetric} log scale to visualize the broad dynamic range. \pp{Comment: Mention why the bottom panel has fewer star numbers than the top panel.}
    Star-level comparison of instantaneous accretion rates at the reference time. Panel (a) compares future massive (FM) stars to non-massive (NM control stars. Only stars with positive instantaneous accretion rates, $\dot{M}_{\rm ref}>0$, are included. Panel (b) shows the same comparison within DBSCAN clusters using the relative accretion rate $\dot{M}_{\rm rel}=\dot{M}_{\rm ref}/\mathrm{median}(\dot{M}_{\rm ref})_{\rm cluster}$, which normalizes each star by the typical accretion level of its host cluster. Panel (b) contains fewer stars because it includes only stars assigned to DBSCAN clusters, unclustered stars are excluded from the within-cluster comparison. Each point represents one star; boxes show the interquartile range, horizontal lines mark the medians, and points are jittered horizontally for visibility. Both panels are shown with logarithmic y-axes.}
    \label{fig:q2_boxplot_column}
\end{figure}

However, a global comparison may still reflect environmental differences, since FM stars preferentially reside in clusters \lng{with more still-accreting stars and higher total stellar masses.} We therefore repeat the comparison after controlling for the host cluster conditions. We normalize each star’s accretion rate by the median accretion rate of its host cluster $\dot{M}_{\rm rel}=\frac{\dot{M}_{\rm ref}}{\mathrm{median}(\dot{M}_{\rm ref})_{\rm cluster}}$. After normalization, the future massive stars still show systematically higher relative accretion: the median value is $\dot{M}_{\rm rel}=1.50$ for future massive stars ($N=47$), compared to $\dot{M}_{\rm rel}=0.76$ for the control sample ($N=191$). The difference remains significant $p=2.5\times10^{-5}$ with a clear shift between the distributions $\delta=0.40$. Figure~\ref{fig:q2_boxplot_column} shows the star-level accretion comparison between future massive and control stars, both globally and after within-cluster normalization This result implies that the enhanced accretion of future massive stars cannot be attributed solely to residing in more active clusters. Future massive stars \pp{have systematically higher accretion rates} than a typical star within their host cluster at the reference time.

\lng{We also apply the same accretion-rate comparison to stars that are not assigned to any DBSCAN cluster at $t_{\rm ref}$. After restricting the sample to stars with $\dot{M}_{\rm ref}>0$, this subset contains 11 FM stars and 26 NM stars.}
Within this subset, FM stars have a higher median accretion rate than NM stars, but the difference is not statistically significant ($p=0.148$, $\delta=0.31$).
Because this comparison is based on a small number of FM stars, we do not draw a strong conclusion about whether the unclustered FM stars have a distinct accretion behaviour.

\subsection{Mass accretion in massive stars}
\label{sec:mass_accretion}

To characterize the growth of future massive stars, we analyze the time evolution of their mass accretion rates, $\dot{M}(t)$. \dana{The accretion rates} \lng{are derived from the time evolution of sink-particle masses.} For each object, we track the stellar mass $M(t)$ from its formation time until it reaches its maximum mass. We estimate the accretion rate as the time derivative of a smoothed mass series, $\dot{M}(t)=dM_{\rm sm}/dt$, \lng{where $M_{\rm sm}(t)$ denotes the smoothed version of the stellar mass history $M(t)$. The smoothing is performed using a Savitzky--Golay filter \citep{SavitzkyGolay1964}, with a window corresponding to $\simeq 0.27\,\mathrm{Myr}$ and a third-order polynomial.}
The analytic derivative yields $\dot{M}$ in units of $M_\odot\, yr^{-1}$. Negative values of $\dot{M}$, which can arise from numerical noise in the discretely sampled mass evolution, are clipped to zero and are not interpreted as physical mass loss.

\begin{figure*}
    \centering
    \includegraphics[width=1\textwidth]{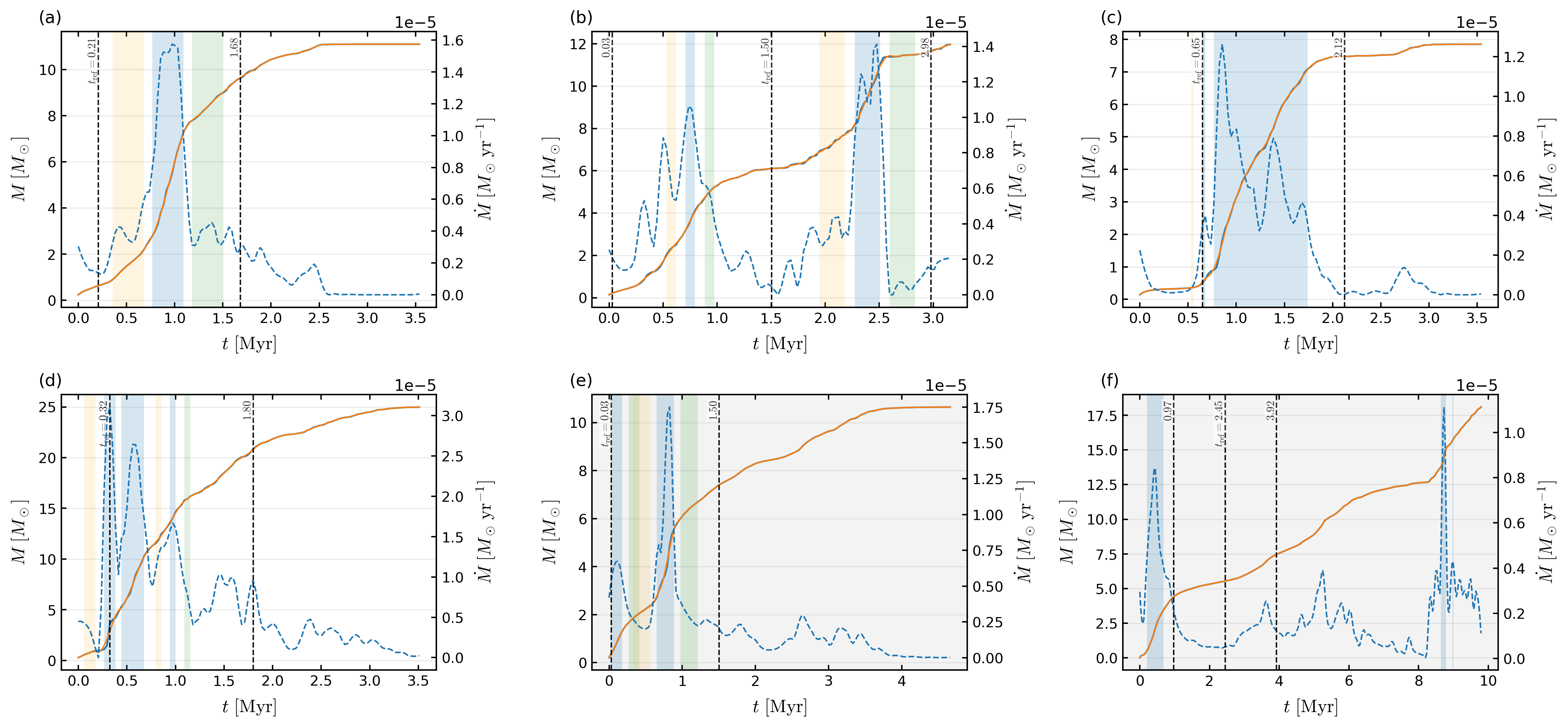}
    % \caption{Overview of the mass growth and accretion-rate variability for the future massive-star sample. For each object, the solid curve shows the \dana{\st{sink}} mass $M(t)$ from formation to the time when it reaches its maximum mass, while the dashed curve shows the accretion rate $\dot{M}(t)$. The shaded intervals mark the enhanced-accretion episodes identified from the accretion-rate time series. Orange and green shaded regions indicate the matched control intervals before and after the enhanced-accretion episode, respectively. Vertical black lines mark $\mathrm{t_{ref}}$ and the boundaries of the analysed time window \dana{\st{. The central line corresponds to $t_{\rm ref}$, while the two outer lines mark the limits of the analysed time window}}, approximately $\pm 1.48$ Myr from $\mathrm{t_{ref}}$ \dana{\st{the reference time}}. The axis limits and numerical ranges are adjusted separately for each star as they span different mass and accretion-rate scales. For readability, axis labels and numeric ranges are suppressed; the figure is intended to illustrate the qualitative diversity of accretion histories and the placement of EA and control intervals across stars. \pp{Comment: we cannot have figure 4 in that form: choose some representative stars (for example illustrating different extremes of behaviour, or three typical cases (single episode, recurrent episodes, broader phases)) and show the other figures in an appendix.}
    \caption{
    Representative accretion histories of selected future massive (FM) stars.
    Panels (a)--(d) show FM stars assigned to DBSCAN clusters at $t_{\rm ref}$, while panels (e)--(f) show FM stars classified as unclustered at $t_{\rm ref}$.
    In each panel, the orange solid curves show the smoothed stellar mass history, $M(t)$, while the dashed blue curve shows the accretion rate, $\dot{M}(t)$, with its scale given on the right axis. \pp{The time $t=0$ corresponds to the formation time of each individual sink particle.}
    Blue-shaded regions mark enhanced accretion (EA) episodes, defined from the per-star accretion-rate history.
    Orange and green shaded regions show the matched pre-EA and post-EA control intervals, respectively, where such intervals are available.
    The central black dashed vertical line marks the reference time, $t_{\rm ref}$.
    The two outer black dashed vertical lines mark the boundaries of the control-search window around $t_{\rm ref}$; they indicate the available time range used to select control intervals and do not correspond to distinct physical events.
    % The examples are selected to illustrate the diversity of FM star growth histories, including single dominant episodes, recurrent episodes, extended phases of enhanced accretion, and high-$M_{\max}$ cases.
}

    \label{fig:3_2_gallery}
\end{figure*}

\lng{Figure~\ref{fig:3_2_gallery} shows representative mass-growth and accretion-rate histories for selected FM stars, including both $M(t)$ and $\dot{M}(t)$. The examples were chosen to illustrate the diversity of accretion behaviour in the FM star sample.}
\lng{The mass histories are generally non-uniform rather than smooth, with rapid growth episodes in $M(t)$ \pp{reflected} by transient peaks in $\dot{M}(t)$.} The accretion variability pattern differs substantially from star to star: some of them exhibit a single dominant early episode followed by reduced accretion, whereas others show recurrent high-accretion intervals or broader phases of increased accretion. 
% \dana{\it{Maybe show examples separately so that the reader can in fact understand these plots? Alternatively, expand Figure 4 over the whole page and get back the x axis.}}

%To quantify the growth stage of future massive stars, we analyze the accretion rate profiles $\dot{M}(t)$. 
We define enhanced accretion (EA) as 
%time intervals 
\dana{episodes} when $\dot{M}(t)$ exceeds a star’s reference level by a fixed factor.
For each star, we define a reference accretion level, $\dot{M}_{60} = P_{60}\big(\dot{M}\big),$
where $P_{60}$ is the 60th percentile of the accretion rate values $\dot{M}(t)$. 
We classify a time step as EA if 
$\dot{M}(t)/\dot{M}_{60} > 2.5$, 
and we require enhanced-accretion intervals to persist for at least two consecutive time steps. Since the time spacing is $\Delta t = 29.5~{\rm kyr}$, this corresponds to a minimum duration of $59~{\rm kyr}$. We adopt $P_{60}$ and the threshold of 2.5 as a feasible compromise between sensitivity and accuracy in identifying EA.

Because individual accretion histories differ strongly from star to star, we use integral metrics that summarize how much mass and how much time are associated with EA.
We characterize each star using (i) the fraction of accreted mass gained during EA, 

\begin{equation}
f_{M,\rm EA}=\frac{\sum_{t\in{\rm EA}}\dot{M}(t)\Delta t}{\sum_{t}\dot{M}(t)\Delta t}
\end{equation}
and (ii) the fraction of time spent in EA, 
\begin{equation}
f_{t,\rm EA}=\frac{\sum_{t\in{\rm EA}}\Delta t}{\sum_{t}\Delta t}.
\end{equation}
For interpretation, we also consider the contrast 
$C=\frac{f_{M,\rm EA}}{f_{t,\rm EA}}\approx \frac{\langle \dot{M}\rangle_{\rm EA}}{\langle \dot{M}\rangle_{\rm all}},$ which measures how much higher the average accretion rate is during EA compared to the time-averaged level.

\begin{figure}
    \centering
    \includegraphics[width=0.48\textwidth]{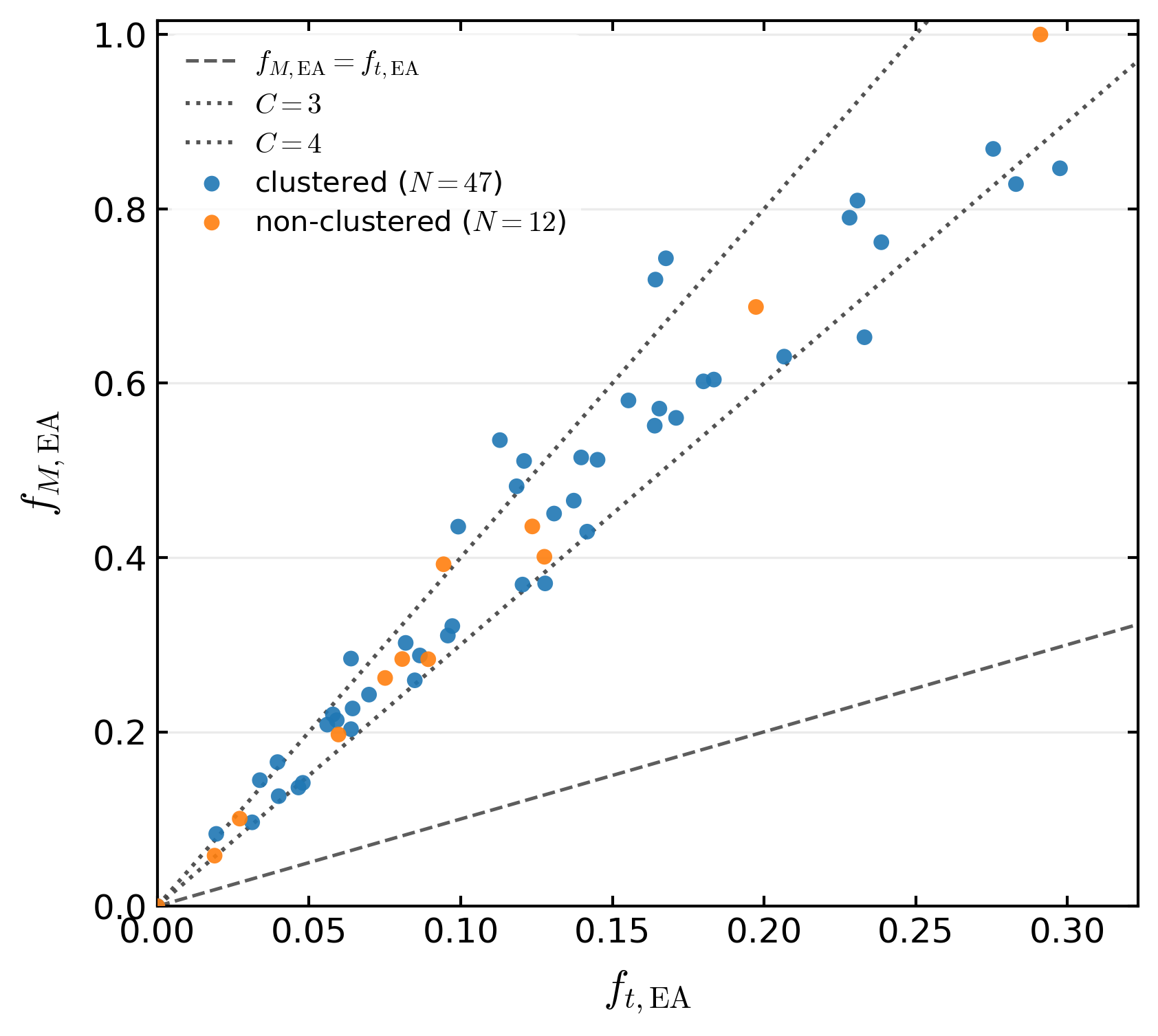}
    \caption{
    Enhanced accretion (EA) time and mass fractions for future massive (FM) stars during their growth intervals. Each point represents one FM star, coloured by its environment at the reference time: clustered stars are shown in blue and unclustered stars in orange. The horizontal axis gives the fraction of the growth time spent in EA, $f_{t,\rm EA}$, and the vertical axis gives the corresponding fraction of mass gained during EA episode, $f_{M,\rm EA}$. The dashed line shows the one-to-one relation, $f_{M,\rm EA}=f_{t,\rm EA}$, while dotted lines show constant accretion-contrast values, $C=f_{M,\rm EA}/f_{t,\rm EA}=3$ and $4$.
    % Mass fraction accreted during enhanced-accretion intervals, $f_{\rm EA}$, versus duty cycle $D$ for all future massive stars (N=59). The  lines $f_{\rm EA}=CD$ correspond to constant contrast of 3 and 4. Points lie predominantly above the diagonal, indicating that enhanced-accretion intervals contribute disproportionately to the total mass growth (median $D=0.113$, median $f_{\rm EA}=0.394$). \pp{Comment: duty cycle $D$ does not appear in section 3.3. Change notation either there or in this figure, for consistency and to avoid confusion.}
    }
    \label{fig:3_2_fEA_vs_D_contrastlines}
\end{figure}

EA episodes are relatively brief but important for mass assembly. In the median case, they occupy only about 10\% of the growth time. Despite this short duration, they contribute a median mass fraction $f_{M,\rm EA}=0.394$ of the \dana{star's total mass.} This means that a small fraction of the accretion history accounts for nearly 40\% of the mass growth. 
The median contrast, $C=3.32$, further shows that the accretion rate during enhanced episodes is typically about three times higher than the time-averaged accretion rate.
\lng{Figure~\ref{fig:3_2_fEA_vs_D_contrastlines} summarizes these results by showing $f_{M,\rm EA}$ versus $f_{t,\rm EA}$, with reference lines of constant contrast $C=f_{M,\rm EA}/f_{t,\rm EA}$, including $C=3$ and $C=4$.}
%In the next section, we use the EA mask, defined by the time steps satisfying the enhanced-accretion criterion, to test whether EA episodes are associated with proximity to junctions in the tracer-derived filament skeleton.%%I copied it to the next section

% \subsubsection{Episodic accretion and the limits of static morphology}

% В РЕЗУЛЬТАТЫ

% We have found that future massive-star growth is strongly episodic. Typically, future massive stars gain about 40\% of their accreted mass during enhanced-accretion intervals that occupy only about 10\% of the growth time. This implies that the connection between future massive-star growth and the surrounding gas morphology should not be inferred from a single reference-time view.
% context
% Instead, the relevant question is whether the gas morphology is connected to the phases when the star is actually growing rapidly. A future massive star may be located near dense gas structure before, during, or after an enhanced-accretion interval, but such proximity at a single time does not by itself show that the structure is relevant to the accretion event. For this reason, we compare enhanced-accretion intervals with control intervals before and after them.

\subsection{Are enhanced-accretion episodes associated with 3D hub proxies?}
\label{sec:3dhubs}
% HFSs are commonly discussed as structures in which filaments may channel material toward a dense central region, or hub. Such hubs are considered possible sites where massive stars can form. In this picture, the geometry of the surrounding gas reservoir may be important for how material is delivered to the forming stars. 

\lng{As discussed in Section~\ref{sec:introduction}, HFSs provide a framework for connecting the growth of future massive stars to the morphology of the surrounding dense gas. Since FM star growth is episodic, we test whether EA episodes are associated with proximity to junctions in the intrinsic morphology at the given time rather than using only a single reference time view. }
%we use the EA mask, defined by the time steps satisfying the enhanced-accretion criterion, to test 

\dana{For this purpose, we reconstructed a 3D skeleton of the gas reservoir traced by particles that are later accreted by stars in each cluster, and used high-connectivity regions of this skeleton as 3D hub proxies. If EA is linked to these 3D hub proxies, EA episodes may be expected to occur when the FM is located closer to them.} 
\dana{The full methodology used to reconstruct the skeletons and identify the 3D hub proxies is presented in Appendix~\ref{app:skeleton}.
We note that these proxies describe the morphology of the dense gas structure, but they are not direct kinematic tracers of gas inflow.
}
%\lng{We test this by measuring the distance between each star and the nearest 3D hub proxy, and by comparing this distance during EA episodes with matched non-EA control intervals before and after them.}

%The full methodology used to reconstruct the skeletons and identify the 3D hub proxies is presented in Appendix~\ref{app:skeleton}.

\begin{figure}
    \centering
    \includegraphics[width=0.38\textwidth]{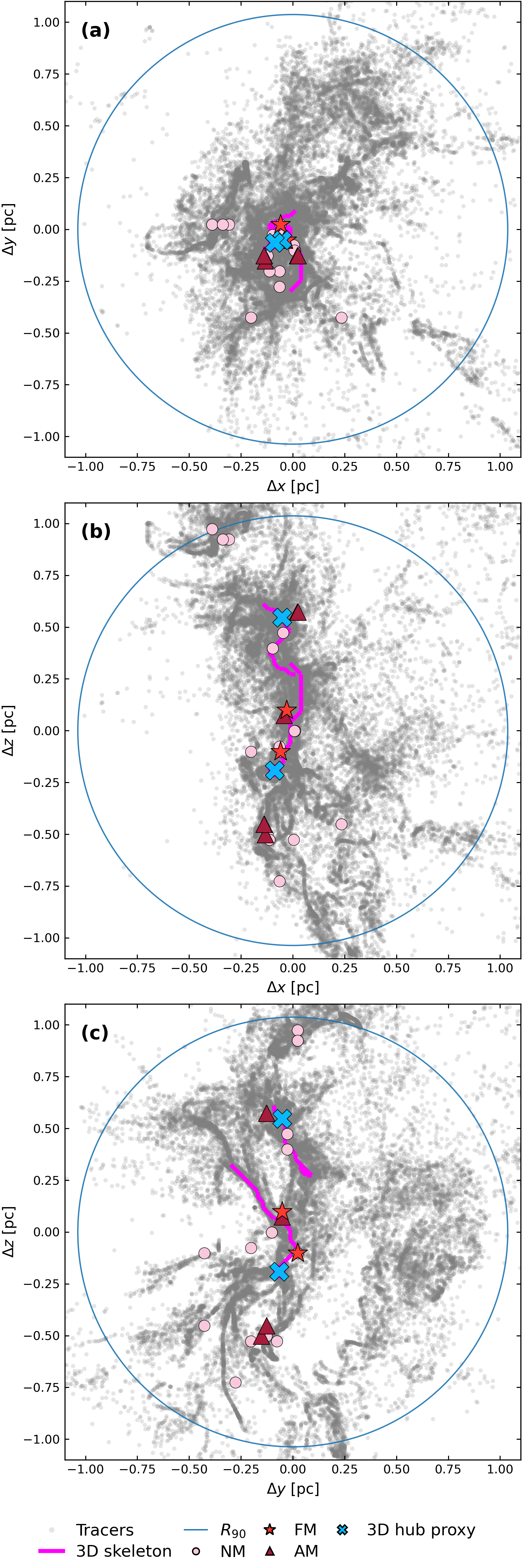}
    \caption{
        Example of the intrinsic cluster geometry of the tracer distribution used to measure distances to 3D hub proxies. The same cluster at $t_{\rm ref}$ is shown in the (a) xy-, (b) xz-, and (c) yz-projections, with coordinates relative to the tracer-defined centre. Grey \dana{dots} show the tracer particles associated with the cluster, while the blue circle marks $R_{90}$, the radius enclosing 90 per cent of the tracer particles in three dimensions. 
        The reconstructed 3D skeleton is shown in magenta, while blue crosses mark the 3D hub proxies. Stellar members are divided into three groups: pale-pink circles show non-massive (NM) stars, bright-red stars show future massive (FM) stars, and burgundy triangles show already massive (AM) stars. 
        % The dashed blue line illustrates the nearest-proxy distance for one representative FM star; in the analysis, this distance is measured in three dimensions and normalized by $R_{90}$.
        % Example visualization of the characteristic size $R_{90}$ \pp{of the tracer distribution} for one cluster at the reference time. Black points show tracer particles associated with the cluster, the red cross marks the geometric center of the tracer distribution, and the blue circle shows $R_{90}$ projected onto the $xy$, $xz$, and $yz$ planes.
        % \lng{Comment: Mika {figure*} - why?}
        }

    \label{fig:r90_example}
\end{figure}

We quantify the spatial relation between each star and the identified 3D hub proxies using the nearest-proxy distance. For each selected cluster within the analysed time window (defined by the time steps satisfying the enhanced-accretion criterion), we compute the 3D distance $d$ from the star position to the nearest junction region. 
At each \dana{time step}, \lng{the nearest 3D hub proxy is defined as the proxy position with the smallest 3D distance to the star.} 
To compare systems with different sizes, we normalize this distance by $R_{90}$. 
\lng{Here, $R_{90}$ is defined using the same tracer particles, i.e. the particles that are later accreted by the stars in the selected cluster. }
\dana{It corresponds to} \lng{the 90th percentile of their 3D distances from the geometric center of the tracer distribution, computed separately for each selected cluster and each time step.} For visualization, Figure~\ref{fig:r90_example} shows an example of this radius projected onto the $xy$, $xz$, and $yz$ planes. The value of $R_{90}$ is computed from the full 3D tracer distribution, while the projected circles are shown only as visual guides. We then use $d/R_{90}$ as the main measure of the relative distance to the nearest junction region. Time steps without a valid junction catalog are not included in this comparison.

To construct the control sample, we match each EA episode to nearby non-EA intervals for the same star. For each EA episode, we define candidate control intervals before and after the EA episode.
Each \lng{retained} control interval is required to have the same duration as the corresponding EA episode. 
% If an EA episode spans $N$ saved outputs, the control intervals before and/or after the episode are selected to span the same number of outputs.
A candidate control interval is retained only if it lies fully within 
%the analysed time window, 
approximately $\pm 1.48$ Myr around $t_{\rm ref}$; otherwise, it is excluded.
% The full control interval must lie within the analysed time window, approximately $\pm 1.48$ Myr around $t_{\rm ref}$. No fallback controls are used: if a pre- or post-control interval does not fit within this window, it is excluded. 
This provides a local-in-time, same-star, and same-duration comparison sample outside the EA episode phase. 
This selection is illustrated in Figure~\ref{fig:3_2_gallery}, where the EA episodes are shown with blue shading. The control windows before and after each EA episode are indicated with orange and green shading, respectively, and in the following we call these the pre-EA and post-EA control samples.

\begin{figure*}
    \centering
    \includegraphics[width=1\textwidth]{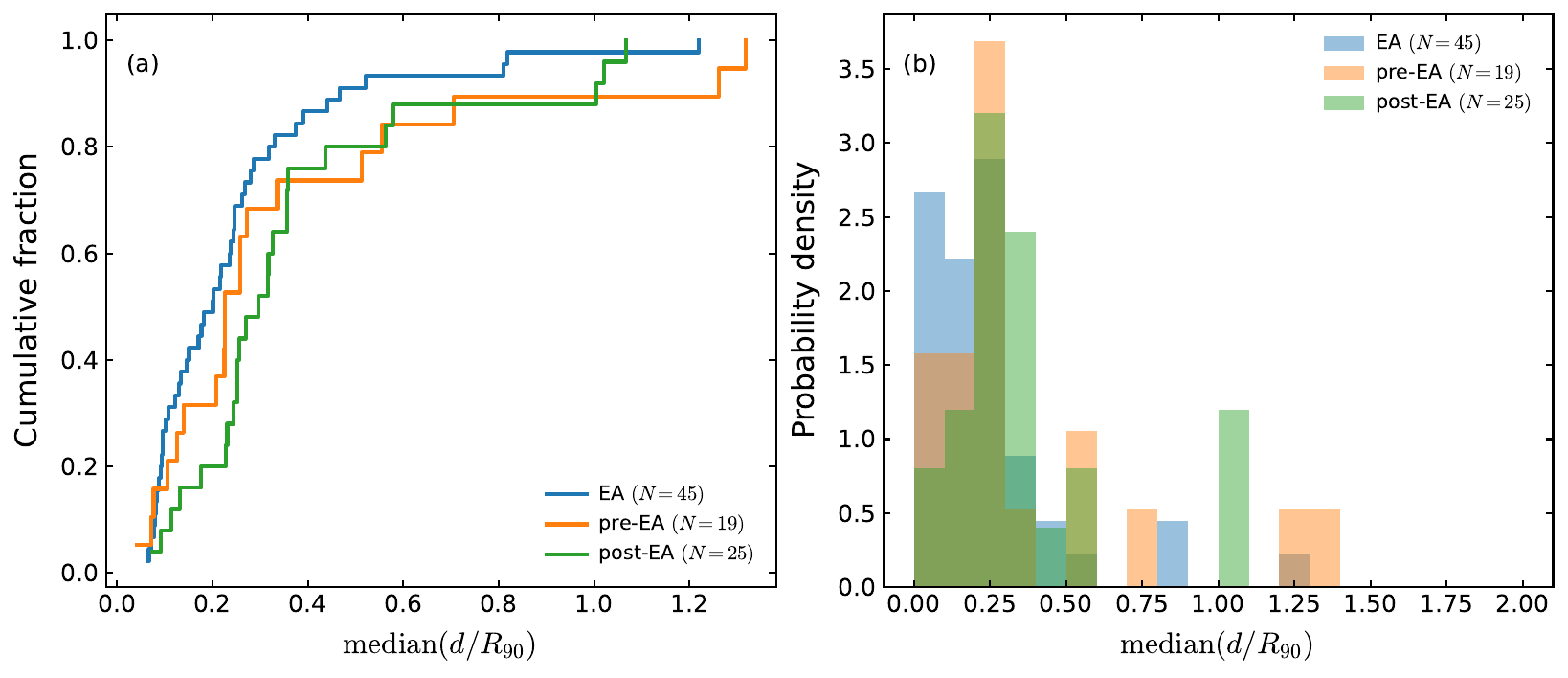}
    \caption{
    Episode-level distributions of the median normalized distance for enhanced accretion (EA) and control episodes. %, $\mathrm{median}(d/R_{90})$,
    Panel (a) shows the cumulative distribution function (CDF), and panel (b) shows the probability-density histogram.
    Blue denotes EA episodes, orange denotes control episodes before EA episodes, and green denotes control episodes after EA episodes.
    The distance $d$ is normalized by $R_{90}$, the 90th-percentile reference radius \dana{of every cluster}.
    % Episode-level distributions of the median normalized distance to the nearest junction region, $\mathrm{median}(d/R_{90})$, for EA episodes, control-before episodes, and control-after episodes. Left: empirical cumulative distribution functions. Right: density histograms with identical binning. \pp{Comment: fix the legend in both panels: EAR --> EA.}
    }
    \label{fig:ear_control_histogram}
\end{figure*}

The resulting episode-level distributions of median $d/R_{90}$ for EA episode and control samples are represented in Figure~\ref{fig:ear_control_histogram}. The cumulative distribution function (CDF) shows, for each value of $d/R_{90}$, the fraction of episodes with distances smaller than or equal to that value. Overall, EA episodes tend to occur at relatively small normalized distances from the nearest junction region, with a large fraction of measurements concentrated at $d/R_{90}\lesssim 1$. In the CDF, this appears as the EA episode curve rising more rapidly at small $d/R_{90}$ values than in the control curves. At the episode level, the median normalized distance to the nearest junction region is 0.20 for EA episodes, compared to 0.23 for the pre-EA control sample and 0.30 for the post-EA control sample. 

\subsection{Projected HFS morphology versus intrinsic 3D structure}
\label{sec:2d3d}

\begin{figure*}
    \centering
    \includegraphics[width=1\textwidth]{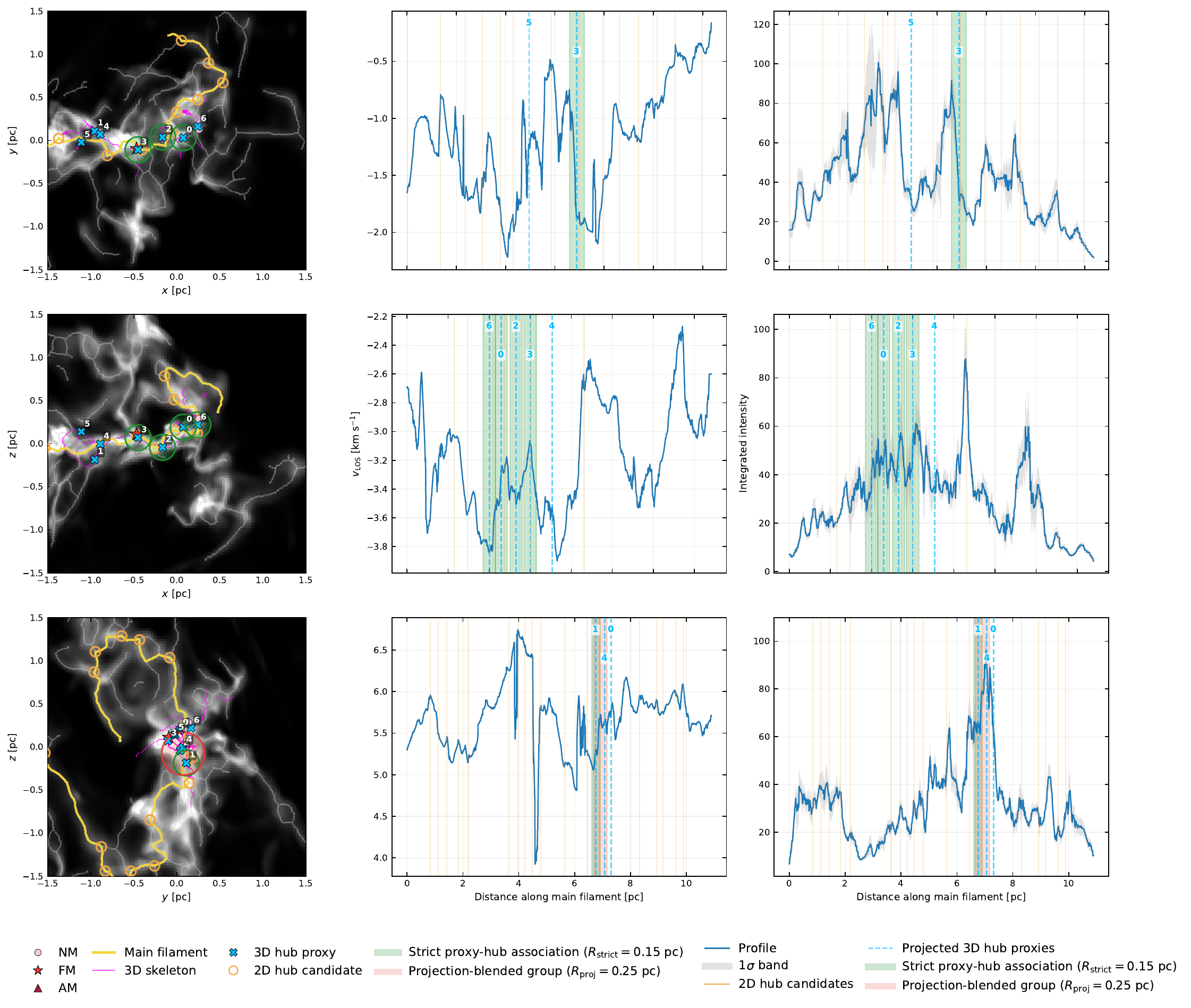}
    \caption{
    Example projected 2D--3D comparison for a cluster at the reference time, $t_{\rm ref}=15.4$ Myr after self-gravity was included \dana{in the simulation}. Rows show the three orthogonal projections: $xy$, $xz$, and $yz$. Left column: synthetic $^{13}{\rm CO}(1$--$0)$ integrated-intensity (moment-0) maps with overlaid stellar positions, projected FilFinder structures, and projected three-dimensional structures. Pink circles, red stars, and dark-red triangles mark non-massive, future massive, and already massive stars at the reference time , respectively. The yellow curve shows the main projected filament identified by FilFinder, orange circles mark 2D hub candidates, magenta line segments show the projected 3D skeleton, and cyan crosses mark projected 3D hub proxies. Middle column: velocity-centroid profiles measured along the main projected filament. Right column: integrated-intensity profiles measured along the same path. Grey shaded regions in the profile panels indicate the $1\sigma$ uncertainty bands. Orange vertical lines mark the positions of 2D hub candidates along the main projected filament, and cyan dashed vertical lines mark projected 3D hub proxies that lie close to the main projected filament. The green and red shaded regions mark selected projected 3D hub-proxy zones shown consistently in the maps and in the corresponding profile panels. Distances are given in pc.
}
    \label{fig:2d_3d_example}
\end{figure*}

\lng{After the intrinsic three-dimensional analysis, we next examine how the same cluster environments appear in projected synthetic molecular-line observations} \dana{and how reliable is the identification of two-dimensional hub candidates with respect to the real morphology}.
\lng{Figure~\ref{fig:2d_3d_example} shows an example of this comparison for one cluster in three orthogonal projections, with the projected 3D hub proxies overlaid on the synthetic $^{13}{\rm CO}$ integrated-intensity maps together with the 2D hub candidates and cluster members.} The numerical labels identify the same intrinsic 3D hub proxies in all three projections, allowing individual proxies to be followed between different viewing directions.
The green circles show the strict association radius, $R_{\rm strict}=0.15\,{\rm pc}$, around matched projected 3D hub proxies, while the corresponding green intervals mark the same regions in the intensity and velocity profiles along the main projected filament. The red circle in the $yz$ projection shows the projected linking radius, $R_{\rm proj}=0.25\,{\rm pc}$, used to highlight a compact group of projected 3D hub proxies. The figure illustrates that the relation between projected hub candidates and intrinsic 3D hub proxies is not the same for different viewing directions.

In the $xy$ projection, the 2D hub candidates, identified from the FilFinder skeleton, are distributed along the main projected filament. While in the next section, we report statistics of the entire sample, here we focus on the region around 3D hub proxy number three, called proxy-3 hereafter.
The proxy-3 region is shown by the green circle in the moment-0 map and by the green shaded interval in the corresponding profile panels. The nearest 2D hub candidate lies within the strict association radius, $R_{\rm strict}=0.15\,{\rm pc}$. 
% \dana{The region is shown as} \lng{the green shaded circle in the moment-0 map and as the green shaded interval in the corresponding profile panels.} 
% \lng{In this projection, the 2D hub candidate identified from the FilFinder skeleton coincides with a projected 3D hub proxy located close to an FM star.} 
This region also corresponds to a clear peak in the position--integrated-intensity ({\ipr}) profile and to a strong variation in the position--velocity ({\vpr}) profile \dana{along the longest filament}, \dana{denoted by the yellow curve in Figure~\ref{fig:2d_3d_example}}. 
% The velocity increases toward this region and then changes again after it, which is the type of behaviour that could be interpreted observationally as a kinematic feature associated with a hub-like structure.
\lng{The velocity-centroid profile increases toward this 2D hub candidate and changes again after it. Similar kinematic signatures near hub regions, including velocity gradients along filaments and V-shaped features in {\vpr} diagrams, have been reported in observational studies of HFSs \citep[e.g.,][]{2022MNRAS.514.6038Z,2025A&A...694L..18B}. We therefore interpret this local change in the velocity-centroid profile as a projected kinematic feature that would be observationally associated with a HFS hub.}
% Although there is a small offset between the projected 2D hub candidate and the 3D hub proxy, this is a favorable example in which the projected moment-0 morphology, the {\vpr} profile, and the {\ipr} profile all identify the same projected hub region. \revi{In this case, proxy 3 is associated with a prominent projected hub feature located close to an FM star, even though the 2D hub candidate and the projected 3D hub proxy are slightly misaligned.} 
Their positions along the main filament are, however, slightly offset from the nearby local features in the {\ipr} and {\vpr} profiles. Nevertheless, the moment-0 morphology and both profiles identify the same broader projected hub region, located close to an FM star.

% In the $xz$ projection, several 2D hub candidates are located close to projected 3D junctions along the main filament. This positional agreement is visible in the profile panels as a group of nearby orange and cyan vertical lines. 
In the $xz$ projection, several 2D hub candidates are located close to projected 3D hub proxies along the main filament. Among them, proxy 3 is particularly informative: it is the same intrinsic 3D hub proxy highlighted in the $xy$ projection, now viewed along a different line of sight. 3D hub proxy-3 is more closely aligned with the 2D hub candidate.
% The intensity and velocity profiles show local variations in the same part of the path, suggesting that the projected structure is not purely morphological. However, these variations are broad and overlapping, so they do not provide a clean one-to-one match for each projected hub candidate or junction. 
The intensity and velocity profiles show local variations in the same part of the path. In particular, the region associated with proxy-3 coincides with a local feature in both the {\ipr} and {\vpr} profiles. In this projection, the projected position of proxy 3 is also closely aligned with the 2D hub candidate, with essentially no visible offset between them and local features in profiles. However, the corresponding profile features are less prominent than in the $xy$ projection, so this case provides a closer positional match but a less visually prominent observational signature.

In the $yz$ projection, the projected 3D junctions are concentrated in a compact part of the main filament. This region coincides with a strong peak in the {\ipr} profile, so it would appear as a clear projected hub in the moment-0 map. However, this apparent agreement should be interpreted with caution, because the intensity peak may be enhanced by projection: several intrinsic junctions overlap in a small area on the sky. The velocity profile in the same region shows local variations, but these variations are embedded in a broader, complex velocity pattern rather than forming a single isolated kinematic signature. Thus, the $yz$ projection illustrates both the strength and the limitation of the projected analysis: an intrinsic concentration of junction regions can appear as a strong intensity feature, but its apparent prominence and interpretation depend on viewing direction and line-of-sight overlap.

Overall, this example shows that \pp{the 2D} morphology may point to physically relevant regions of the intrinsic \pp{3D} structure, but the interpretation is not always direct. A projected hub candidate may coincide with a \pp{3D} hub proxy and with clear intensity or velocity features. However, a strong intensity peak is not automatically a unique physical hub, because it can be enhanced by the projection of several intrinsic structures along the line of sight. Besides, when the {\vpr} and {\ipr} profiles are broad or complex, the projected hub candidate is more difficult to interpret. Such profiles do not necessarily rule out an underlying \pp{3D} hub structure, but they may indicate that several components are blended in projection. \pp{PI and PV} profiles are therefore important diagnostics and should be used together with moment-0 morphology when interpreting projected hub candidates. They help test whether a projected hub feature is also associated with coherent intensity and velocity structure. However, these profiles are still based on projected data, so they cannot guarantee a fully accurate identification of the underlying 3D structure. 
% The example in Figure~\ref{fig:2d_3d_example} illustrates both favourable and ambiguous cases. In one projection, a 2D hub candidate can coincide with a projected 3D hub proxy and with clear intensity and velocity-centroid features. In another projection, several intrinsic hub-proxy regions can overlap on the sky and produce a stronger, but less uniquely interpretable, projected hub feature. We therefore next quantify this effect statistically over the full sample of projected 3D hub proxies.
The example in Figure~\ref{fig:2d_3d_example} illustrates several types of correspondence. In the $xy$ projection, a 2D hub candidate and projected 3D hub proxy 3 are closely associated, while their positions are slightly offset from prominent local features in the {\ipr} and {\vpr} profiles. In the $xz$ projection, the same proxy shows a closer positional match with the 2D hub candidate and coincides with local features in both profiles, although these features are less pronounced. In the $yz$ projection, several intrinsic 3D hub proxies overlap on the sky and produce a prominent, but less uniquely interpretable, projected hub feature. We therefore next quantify these effects statistically over the full sample of projected 3D hub proxies.

\subsection{Statistical recovery of 3D hub proxies}
For each 3D hub proxy, we measure the projected distance to the nearest 2D hub candidate in each of the three orthogonal projections. This gives 333 projected hub-proxy appearances, corresponding to 111 physical 3D hub proxies viewed along three lines of sight. We define a projected recovery when the proxy--hub distance is below an adopted association radius, using two operational thresholds: $R_{\rm strict}=0.15\,{\rm pc}$ for close matches and $R_{\rm loose}=0.25\,{\rm pc}$ for more loose matches. The strict association radius is illustrated in Figure~\ref{fig:2d_3d_example} by the green circles centred on selected projected 3D hub proxies. Across all projected appearances, the strict criterion recovers 91 out of 333 projected 3D hub-proxy positions, corresponding to $27.3\%$. With the loose criterion, the recovery increases to 162 out of 333, or $48.6\%$. Thus, a 3D hub proxy that is present in the intrinsic structure is not necessarily recovered as a 2D hub candidate in a single projected view.
% We adopt two projected association radii. The strict radius,
% $R_{\rm strict}=0.15\,{\rm pc}$, is close to the 25th percentile of the
% projected 3D-proxy--2D-hub distance distribution and lies just beyond
% the small-separation peak. The loose radius,
% $R_{\rm loose}=0.25\,{\rm pc}$, is close to the median distance and is
% also consistent with a shallow secondary minimum around
% $0.23{-}0.25\,{\rm pc}$ in the finely binned histograms. We use these
% two thresholds to distinguish close projected coincidences from a more
% permissive projected association. 

\begin{table}
\centering
\caption{
Recovery of physical 3D hub proxies across viewing directions. 
For each 3D hub proxy, we count in how many of the three projections it is associated with a 2D hub candidate. 
The strict and loose criteria use $R_{\rm match}=0.15\,{\rm pc}$ and $0.25\,{\rm pc}$, respectively.
}
\label{tab:proxy_recovery_nproj}
\footnotesize
\setlength{\tabcolsep}{3.5pt}
\begin{tabular}{lcccc}
\hline
Criterion & $N_{\rm view}=0$ & $N_{\rm view}=1$ & $N_{\rm view}=2$ & $N_{\rm view}=3$ \\
\hline
Strict & 47 (42.3\%) & 41 (36.9\%) & 19 (17.1\%) & 4  (3.6\%) \\
Loose  & 17 (15.3\%) & 44 (39.6\%) & 32 (28.8\%) & 18 (16.2\%) \\
\hline
\end{tabular}
\end{table}

\rev{Table~\ref{tab:proxy_recovery_nproj} shows the recovery of the same physical 3D hub proxies across the three viewing directions. With the strict radius, 64 out of 111 3D hub proxies are recovered in at least one projection, but only 23 are recovered in two or three projections, and only 4 are recovered in all three projections. With the loose radius, 94 out of 111 proxies are recovered in at least one projection, but only 18 are recovered in all three projections. Therefore, the recovery of 3D hub proxies in projected maps is not only incomplete, but also strongly dependent on viewing direction.}

% \begin{table}
% \centering
% \caption{
% Stars inside matched 3D hub-proxy regions. Counts are summed over all matched 
% projected regions. The strict and loose radii are $0.15\,{\rm pc}$ and 
% $0.25\,{\rm pc}$, respectively.
% }
% \label{tab:stars_inside_matched_regions}
% \begin{tabular}{lccccc}
% \hline
% Criterion & $N_{\rm regions}$ & $N_{\rm AM}$ & $N_{\rm FM}$ & $N_{\rm NM}$ & $N_{\rm total}$ \\
% \hline
% Strict & 91  & 3  & 42  & 75  & 120 \\
% Loose  & 162 & 13 & 105 & 237 & 355 \\
% \hline
% \end{tabular}
% \end{table}

% \begin{table}
% \centering
% \caption{
% Projection multiplicity of star--proxy associations. For each star--proxy pair, 
% we count in how many projections the star falls inside the matched projected 
% hub-proxy region.
% }
% \label{tab:star_proxy_projection_multiplicity}
% \begin{tabular}{lcccc}
% \hline
% Criterion & $N_{\rm pairs}$ & 1 projection & 2 projections & 3 projections \\
% \hline
% Strict & 88  & 60 (0.682)  & 24 (0.273) & 4  (0.045) \\
% Loose  & 242 & 159 (0.657) & 53 (0.219) & 30 (0.124) \\
% \hline
% \end{tabular}
% \end{table}

\rev{As a secondary check, we count the stars that fall within the matched projected hub-proxy regions. These are association counts rather than unique-star counts, because the same star--proxy pair can be detected in more than one projection. For the strict radius, the 91 matched projected regions contain 120 star--region associations, including 42 FM, 75 NM, and 3 AM associations. For the loose radius, the number increases to 355 associations, including 105 FM, 237 NM, and 13 AM associations. Most star--proxy associations are visible in only one projection: 60 out of 88 pairs for the strict radius and 159 out of 242 pairs for the loose radius. This indicates that projected star--hub associations are also sensitive to viewing direction.}

% \begin{tabular}{lcccc}
% \hline
% Criterion & $R_{\rm 3D,sep}$ & $N_{\rm groups}$ & $N_{\rm clusters}$ & Median max 3D sep [pc] \\
% \hline
% Strict & 0.30 pc & 9  & 6  & 0.409 \\
% Strict & 0.50 pc & 4  & 3  & 0.782 \\
% Loose  & 0.30 pc & 26 & 11 & 0.443 \\
% Loose  & 0.50 pc & 11 & 8  & 0.894 \\
% \hline
% \end{tabular}

\rev{
In a projected observation, a 2D hub candidate does not necessarily correspond to a single intrinsic 3D hub proxy, because several 3D proxies can overlap along the line of sight. We therefore test how often several 3D hub proxies appear close together in one projection, but are separated when viewed from the other directions. For each cluster and projection, we apply DBSCAN to the projected positions of the 3D hub proxies to identify compact projected groups. We classify a compact projected group as a projection-blended candidate if its members are spatially separated in at least one of the other two orthogonal projections. For the conservative case, we use a DBSCAN linking length of $R_{\rm proj}=0.15\,{\rm pc}$ in the reference projection and require the maximum pairwise separation in the other projections to exceed $R_{\rm other}=0.50\,{\rm pc}$. With this conservative criterion, we identify 15 projection-blended groups in 11 projected maps and 8 clusters. The median maximum separation in the other projections is $0.72\,{\rm pc}$, indicating that these compact projected groups do not correspond to compact 3D structures. Using a looser projected linking length, $R_{\rm proj}=0.25\,{\rm pc}$, increases the number of candidates to 22 in 18 projected maps and 10 clusters. The looser projected linking scale is illustrated in Figure~\ref{fig:2d_3d_example} by the red circle in the $yz$ projection, which highlights a compact projected group of 3D hub proxies identified using $R_{\rm proj}$. The corresponding red shaded intervals in the second and third columns mark the same region along the main projected filament in the profile panels.} Thus, some compact projected groups associated with 2D hub candidates can represent line-of-sight blends of multiple intrinsic 3D hub proxies, rather than a single compact 3D structure.

\rev{Overall, these results show two complementary projection effects. First, intrinsic 3D hub proxies are not always recovered as 2D hub candidates in a single projected view. Second, some compact projected groups can combine multiple intrinsic 3D hub proxies along the line of sight. Therefore, projected HFS morphology can trace physically relevant parts of the intrinsic dense-gas structure, but the correspondence between 2D hub candidates and 3D hub proxies is incomplete, viewing-direction dependent, and sometimes affected by line-of-sight blending.}

\section{Discussion}
\label{sec:discussion}

\subsection{The environmental context of future massive-star growth}

\rev{In Section~3.1, we examined the local stellar environments of still-accreting stars at the reference time. We compared \pp{stellar} groups with \pp{FM} stars to those without \pp{them}.} We found that \pp{FM} stars are preferentially associated with larger and more massive stellar groups at the reference time. These groups contain more forming stars and have higher total stellar mass than groups without \pp{FM} stars. This result indicates that \pp{FM}-star growth is not randomly distributed among the forming stellar population, but is preferentially embedded in larger and more active local star-forming environments. \rev{This environmental classification is defined at one reference time, whereas stellar growth occurs over an extended accretion history. Thus, this association should be interpreted as a reference-time view of an evolving star-forming environment.} \rev{The preferential association of \pp{FM} stars with larger and more active stellar environments is} consistent with the broader view that massive-star formation is connected to dense and clustered regions \citep{2003ARA&A..41...57L, 2007ARA&A..45..481Z, 2018ARA&A..56...41M,2026A&A...705A..79A}.

% In the present simulation, this environment is not static: the gas evolves within a turbulent, feedback-regulated ISM, where local reservoirs can be dispersed or reorganized over time \citep{2020SSRv..216...50C, 2020ApJ...900...82P, Padoan2016, Haugbolle2018}. The association with larger stellar groups therefore suggests that future massive stars can still assemble mass within structured, gas-rich environments during this dynamical evolution. A larger stellar group may indicate a larger local gas reservoir, stronger local collapse, a deeper gravitational potential, or a more developed star-forming region \citep{2003ApJ...585..850M, 2004MNRAS.349..735B, 2014prpl.conf..149T, 2019MNRAS.490.3061V, 2020ApJ...900...82P}. In this sense, the group properties are not only descriptive cluster statistics, but environmental diagnostics of the conditions in which future massive stars continue to assemble their mass \citep{2014prpl.conf..149T, 2018ARA&A..56...41M, 2020ApJ...900...82P}. 

\rev{Previous studies provide several ways to interpret the role of the local environment in massive-star formation. In turbulent-core or core-accretion models, the dense \pp{prestellar } core \pp{provides} the \pp{full} reservoir for massive-star growth \citep{McKeeTan2002,2003ApJ...585..850M}. In competitive-accretion and cluster-scale accretion models, the shared gas reservoir and the gravitational potential of the forming stellar group play a more direct role in regulating stellar mass growth \citep{2004MNRAS.349..735B, 2006MNRAS.370..488B, 2014prpl.conf..149T}. More global pictures of star formation emphasize hierarchical collapse, turbulent flows, and continued accretion from larger-scale gas reservoirs \citep{2019MNRAS.490.3061V, 2020ApJ...900...82P,2026A&A...705A..79A}.}

In the simulation used in this study, this environment is not static: the gas evolves within a turbulent, feedback-regulated ISM, where local reservoirs can be dispersed or reorganized over time \citep{Padoan_2016,2020SSRv..216...50C, 2020ApJ...900...82P, Lu+2020}. \rev{Our result adds to this broader picture by showing that future massive stars are preferentially associated with larger stellar groups even within this dynamically evolving environment.} The association with larger stellar groups therefore suggests that future massive stars can still assemble mass within structured, gas-rich environments during this dynamical evolution. \rev{In our analysis, the larger stellar group is not treated as a direct mass-delivery mechanism. Rather, it may trace a larger local gas reservoir, stronger local collapse, a deeper gravitational potential, or a more developed star-forming region.} In this sense, the group properties are not only descriptive cluster statistics, but environmental diagnostics of the conditions in which future massive stars continue to assemble their mass.

\rev{The interpretation of the stellar environment at the reference time also depends on the definition of the working sample.} The working sample is restricted to stars that are still in a significant growth phase at the reference time, with $M_{\rm ref} < 0.9 M_{\rm max}$. Within this sample, stars may already be above the massive star threshold or may still be below it at the reference time, and the future massive label is assigned from their later maximum mass. The stellar group identified at the reference time may therefore trace a local star forming environment and gas reservoir that continues to evolve during the subsequent growth of the future massive star. \rev{Therefore, the cluster association should be treated as a time-dependent description of the local environment, not as a fixed category.}

\rev{The unclustered category provides a related caveat and also illustrates the time-dependent nature of the environment.} Most future massive stars that are unclustered at the reference time become associated with stellar groups later. Thus, the stellar environment provides important context for future massive-star growth, but understanding the growth process requires connecting this environment to time-dependent accretion histories and to the evolving gas morphology.

\subsection{Link between 3D hub proxies and enhanced-accretion episodes}

\rev{In Section~\ref{sec:3dhubs}, we compared EA (enhanced-accretion) episodes matched control intervals using the distance between future massive stars and 3D hub proxies in the tracer-defined dense gas.} 
The clearest contrast is seen relative to the post-EA control. Meanwhile, the pre-EA control is more similar to the EA episode distribution, but with a weaker concentration at small distances and a more extended tail toward larger values. 
Taken together, this shows that EA phases tend to occur when future massive stars are close to 3D hub proxies.
%We found that EA episodes occur slightly closer to  than the control intervals. 
\rev{This result suggests that enhanced accretion is not randomly distributed with respect to the intrinsic dense-gas morphology, but is preferentially associated with regions where dense gas structures converge.}
% Thus, the result suggests that enhanced accretion is preferentially associated with the converging parts of the intrinsic dense-gas morphology, rather than being randomly distributed with respect to the surrounding gas structure.
 
We focus on 3D hub proxies rather than on complete HFS identification because HFSs can have diverse morphologies and ambiguous boundaries \citep{2022MNRAS.514.6038Z}. Converging dense regions are the common feature shared by HFSs, even when the full filamentary configuration differs from system to system \citep{2020A&A...642A..87K, 2022MNRAS.514.6038Z}. \rev{Thus, we use 3D hub proxies as operational markers of the local dense-gas morphology around growing stars.}

\rev{Previous observational and theoretical studies often interpret HFSs as multi-scale gas-supply structures, in which filaments connect larger gas reservoirs to dense central hubs where material can accumulate and clustered or massive star formation can proceed \citep{2019A&A...629A..81T, 2020A&A...642A..87K}. Broader theoretical scenarios also emphasize that accretion may continue to be supplied from outside the immediate star-forming core, through converging flows or local dense structures within a larger reservoir \citep{2019MNRAS.490.3061V, 2020ApJ...900...82P}.}
\dana{Recent simulations of centrally concentrated molecular clouds similarly show that global collapse drives gas toward the cluster center, producing compact central clusters and concentrating massive-star formation within the densest regions of the cloud \citep{2026A&A...705A..79A}.}
\rev{Our result adds a time-dependent accretion perspective to this picture by showing that FM stars are located closer to intrinsic 3D hub proxies during EA episodes than during the matched control intervals.}
\rev{Within this framework, the 3D hub proxy is best understood as a marker of the larger-scale converging reservoir, not as the direct source of all accreted material.} Local dense gas concentrations within this reservoir may be more closely connected to rapid stellar growth.

However, the proximity trend should not be interpreted as direct evidence for gas delivery by itself. The 3D hub proxies used here are morphological indicators: they identify where dense gas structures meet, but they do not directly measure velocity convergence, mass flux, or gas inflow toward the star. Our result supports a time-dependent spatial association between enhanced accretion and intrinsic HFS morphology, but establishing a direct mass-delivery mechanism requires kinematic information in addition to morphology \citep{2019A&A...629A..81T, 2019ApJ...875...24C, 2022ApJ...931..115W, 2024A&A...688A..86Z}.
We therefore treat this comparison as evidence for a possible spatial association between enhanced accretion and junction geometry, while leaving the physical origin of this association open \pp{to} further investigation.

%However, this result should be interpreted cautiously. The comparison is based only on episodes for which a valid junction catalog could be constructed, and the number of usable EA episodes is limited after quality control. In addition, the identified junction regions are operational skeleton features derived from the tracer distribution, rather than direct observational or kinematic evidence of gas inflow. 
%We therefore treat this comparison as evidence for a possible spatial association between enhanced accretion and junction geometry, while leaving the physical origin of this association open \ppst{for} \pp{to} further investigation.

\subsection{From 3D hub proxies to 2D hub candidates}
In Section~\ref{sec:2d3d}, we used synthetic molecular-line observations to compare projected 2D hub candidates with the projected positions of intrinsic 3D hub proxies. We have found that this 3D--2D correspondence depends strongly on viewing direction. In favourable projections, a 2D hub candidate can coincide with the projected position of a 3D hub proxy and can also be associated with clear intensity or velocity features along the projected filament. In other projections, however, the correspondence is less direct, because several intrinsic dense structures can overlap along the line of sight.

\rev{Because the 3D--2D correspondence depends on viewing direction,} a projected hub candidate should not be interpreted automatically as a unique physical 3D hub.
\rev{The viewing direction dependence found here is consistent with previous simulation-based studies showing that projection can distort the inferred properties and physical interpretation of molecular-cloud structures. \citet{2010ApJ...712.1049S} showed that projection can change the derived mass--size and linewidth--size relations of cloud structures, while \citet{2013ApJ...777..173B} quantified how superposition in synthetic molecular-line observations introduces uncertainties in cloud properties derived from PPV data. \citet{2019MNRAS.485.4509L} further showed that projection can lead to misleading interpretations of the true three-dimensional shape, size, and velocity structure of filamentary molecular clouds. Similarly, \citet{2012A&A...544A.141J} showed that line-of-sight confusion in synthetic sub-millimetre continuum observations can create apparent filamentary structures, further illustrating that projected morphology may not map uniquely onto the intrinsic three-dimensional gas structure.}

\rev{Our results extend this projection issue to the identification of projected hub candidates:} a strong integrated-intensity peak may mark a physically relevant region, but it may also be enhanced by projection when several intrinsic structures overlap on the sky. 
Similarly, broad or complex {\ipr} and {\vpr} profiles do not necessarily exclude the presence of an underlying 3D hub structure. \rev{However, they make the correspondence between the projected feature and the intrinsic 3D structure less unique, because multiple density and velocity components can contribute to the same projected signal.}

Therefore, 2D hub candidates should be treated as observational hypotheses rather than as direct identifications of intrinsic 3D hub proxies. Moment-0 or column density morphology are useful for identifying candidate regions, and {\ipr} and {\vpr} diagnostics provide an additional test of whether the projected feature is associated with coherent intensity or velocity structure \citep{2019A&A...629A..81T, 2019ApJ...875...24C, 2022ApJ...931..115W}. However, these diagnostics are still based on projected data and cannot fully remove line-of-sight ambiguity \citep{2013ApJ...777..173B, 2019MNRAS.485.4509L}. This reinforces the need to interpret observed HFS morphology with projection effects in mind.

\section{summary and Conclusions}
\label{sec:conclusion}
Despite the fundamental role of massive stars in the evolution of their environments, the processes that control their early mass growth remain poorly understood. As this growth occurs within dense, structured, and often clustered gas, the morphology of the surrounding material may provide important clues to how massive stars assemble their mass. HFSs are a promising framework for studying this connection, because hubs are thought to be the sites where massive stars form, while filaments may supply them with gas. However, it remains unclear how exactly the early accretion of future massive stars is connected to their clustered environment and to the hub-like geometry of the surrounding gas. This connection is even more difficult to assess in \pp{PPV} data, where projection effects and line-of-sight confusion can alter the apparent filamentary structure. In this work, we examined the early growth of future massive stars in the context of HFSs in simulation data, focusing on how their accretion is related to clustered environments and to junction regions in the surrounding dense gas morphology. We also explored how these structures appear in synthetic molecular-line observations, and how the intrinsic three-dimensional morphology is seen in projected \pp{PPV} data.

We approached this question from the stellar population to the surrounding gas structure and then to its observational appearance. First, we selected the still-accreting stellar population from the sink-particle histories and identified the subset of stars that will later become massive. We then used DBSCAN to characterize their clustered environments. Next, we reconstructed time-dependent accretion histories in order to identify enhanced-accretion intervals. To test whether these intervals are related to hub gas geometry, we built tracer-based three-dimensional dense structures, extracted their skeletons, and identified junction branch regions. Finally, using LOC synthetic molecular-line observations, we analysed projected moment maps to examine how the intrinsic three-dimensional morphology appears in observations.

The main results of our study are the following:
\begin{enumerate}
    \item Future massive stars gain about 40\% of their accreted mass during enhanced-accretion intervals that occupy only about 10\% of their growth time. In addition, these enhanced-accretion intervals occur preferentially when future massive stars are close to junction regions of the tracer-derived dense-gas morphology, which we use as proxies for HFS. At the episode level, the median normalized distance is $d/R_{90}=0.20$ during enhanced accretion, compared to 0.23 for the pre-EA sample and 0.30 for the post-EA sample.
    \item Projected moment maps can trace some signatures of the intrinsic three-dimensional structure, but the interpretation is projection-dependent. In favorable cases, projected hub candidates coincide with projected three-dimensional junction regions, future massive stars, and coherent intensity or velocity features. At the same time, strong intensity peaks can be enhanced by line-of-sight overlap, and {\ipr} and {\vpr} profiles may remain complex. Therefore, position--intensity and position--velocity diagnostics should be used together with 2D morphology when interpreting hub structures in observations, although they cannot fully remove the ambiguity caused by projection effects.
\end{enumerate}

Overall, this work supports a picture in which the early growth of future massive stars is linked both to their clustered stellar environment and to the dense, hub-filament morphology of their surrounding gas reservoir. Future massive stars are preferentially found in clusters with more forming stars, show stronger accretion than lower-mass control stars, and assemble a substantial fraction of their mass during short enhanced-accretion episodes. The comparison between enhanced-accretion episodes and same-star control intervals provides tentative evidence for the impact of HFS on the early growth of massive stars. At the same time, connecting this intrinsic three-dimensional picture to observations requires careful treatment of projection effects and synthetic-observation diagnostics.

Taken together, our results suggest that future massive-star growth is connected to the surrounding environment in a time-dependent way. Future massive stars preferentially occur in larger forming stellar groups, but their growth is not continuous. Moreover, HFS morphology should not be interpreted only as a static location where massive stars form. Instead, it may represent an evolving gas-reservoir geometry that is more relevant during particular phases of rapid growth. In this interpretation, the stellar group provides the broader environmental context, while the 3D hub proxies trace local dense-gas morphology that may be associated with enhanced accretion.

At the same time, several limitations should be kept in mind. The stellar grouping analysis does not identify a mass-delivery mechanism by itself. The 3D hub proxies are morphological indicators rather than direct measurements of inflow, and the synthetic observations show that projected 2D hub candidates can be affected by viewing direction and line-of-sight overlap. Therefore, the connection between HFS morphology and future massive-star growth should be interpreted as evidence for a time-dependent spatial association, not as proof of direct gas delivery.

Future work should test this physical picture with kinematic information. In particular, tracer velocities, velocity convergence, and mass-flux estimates are needed to determine whether the 3D hub proxies are also sites of gas inflow toward future massive stars. A more systematic comparison across different viewing directions and a time-dependent tracking of stellar groups would also help determine how reliably projected HFS candidates recover the intrinsic three-dimensional gas morphology.

\section*{Acknowledgements}
This work was supported by the Faculty Development Competitive Research Grant Program of Nazarbayev University No. 201223FD8821.
\pp{PP acknowledges support by the US National Science Foundation under Grant AST 2408023.} MJ acknowledges the support of the Research Council of Finland Grant No. 348342. NKB acknowledges the support of the China Postdoctoral Science Foundation through grant No. 2025M773187. 

%%%%%%%%%%%%%%%%%%%%%%%%%%%%%%%%%%%%%%%%%%%%%%%%%%
\section*{Data Availability}
The synthetic observations generated in this study will be made publicly available through the Kazakhstan National Virtual Observatory (KazVO) upon publication of this article.



\bibliographystyle{mnras}
\bibliography{example} % if your bibtex file is called example.bib

%\begin{thebibliography}{99}
%\bibitem[\protect\citepauthoryear{Author}{2012}]{Author2012}
%\bibitem[\protect\citepauthoryear{Others}{2013}]{Others2013}
%\end{thebibliography}

%%%%%%%%%%%%%%%%%%%%%%%%%%%%%%%%%%%%%%%%%%%%%%%%%%

%%%%%%%%%%%%%%%%% APPENDICES %%%%%%%%%%%%%%%%%%%%%

\appendix

\section{Skeletonization and identification of junction regions}
\label{app:skeleton}

For each clustered group of stars, we load the tracer particles that will be accreted by those stars in the future, in order to identify the surrounding gas reservoir. The resulting tracer distribution represents the three-dimensional spatial structure of the material associated with the group. To reconstruct its dense backbone, we first compute the geometric center of the tracer distribution, recenter the tracer coordinates, and restrict the analysis to a fixed three-dimensional box. Within this volume, we estimate the local tracer density using voxel counts and select only the dense tracers above a chosen threshold. In this way, the analysis is focused on the most populated part of the tracer distribution rather than on sparse outer particles.

The selected dense tracers are then converted into a voxelized density field on a regular grid. To reduce noise and connect nearby dense features, we smooth this field with a Gaussian kernel and then apply a threshold to construct a binary mask. Basic morphology cleaning is further applied in order to remove small isolated structures and obtain a more continuous filamentary volume. We then identify the connected components of the binary mask, rank them by size, and retain the top components for further inspection. The largest component is used as the main structure in the primary analysis, while lower-rank components are also saved as diagnostic products.

Next, we compute a three-dimensional skeleton from the cleaned mask, reducing the volume to a one-voxel-thick centerline representation of the dense filamentary network. For each retained component, the skeleton is converted into a voxel graph. We then classify the skeleton nodes according to their local connectivity, by counting how many neighboring skeleton elements are connected to a given node. Nodes with one connection are treated as endpoints, nodes with two connections form chain segments, and nodes with three or more connections are treated as branch nodes. In order to reduce small irregular features produced by discretization or mask roughness, we prune short terminal branches while preserving the main topology of the structure. The graph is then simplified by collapsing degree-two chains into single edges.

It should be noted that branching in the skeleton often appears not as a single voxel, but as a small connected group of nearby junctions. To avoid assigning several junctions to the same local branching area, we merge them that are connected by short edges and group them into branch regions. Each branch region is then represented by a single node at its centroid. These branch-region nodes provide a practical way to identify candidate three-dimensional junction locations, because they mark where several skeletal branches meet. We treat them as operational tracers of possible hub positions, while noting that their exact number and location may still depend on the adopted reconstruction parameters.

The final set of branch regions is used for analysis and visualization and serves as a proxy for the branching topology of the dense gas reservoir that will later be accreted by the target stars.

%%%%%%%%%%%%%%%%%%%%%%%%%%%%%%%%%%%%%%%%%%%%%%%%%%

% Don't change these lines
\bsp	% typesetting comment
\label{lastpage}
\end{document}